\documentstyle[12pt]{article}

\def\a{\begin{eqnarray}}
\def\b{\end{eqnarray}}
\def\0{\nonumber}
\input amssym.def
\font\teneusm=eusm10                    
\font\seveneusm=eusm7                   
\font\fiveeusm=eusm5                    
\newfam\eusmfam
\textfont\eusmfam=\teneusm
\scriptfont\eusmfam=\seveneusm
\scriptscriptfont\eusmfam=\fiveeusm

\def\dg{ dressing group}

\def\dges{ dressing group elements}

\renewcommand{\theequation}{\thesection.\arabic{equation}}

\newlength{\extraspace}
\newlength{\extraspaces}
\newcounter{dummy}

\newcommand{\ai}{
\addtocounter{equation}{1}
\setcounter{dummy}{\value{equation}}
\setcounter{equation}{0}
\renewcommand{\theequation}{\thesection.\arabic{dummy}\alph{equation}}
\begin{eqnarray}
\addtolength{\abovedisplayskip}{\extraspaces}
\addtolength{\belowdisplayskip}{\extraspaces}
\addtolength{\abovedisplayshortskip}{\extraspace}
\addtolength{\belowdisplayshortskip}{\extraspace}}
\newcommand{\bj}{
\end{eqnarray}
\setcounter{equation}{\value{dummy}}
\renewcommand{\theequation}{\thesection.\arabic{equation}}}

\input amssym
\newcommand{\be}{\begin{equation}}
\newcommand{\ee}{\end{equation}}
\newcommand{\ba}{\begin{eqnarray}}
\newcommand{\ea}{\end{eqnarray}}
\newcommand{\ban}{\begin{eqnarray*}}
\newcommand{\ean}{\end{eqnarray*}}
\newcommand{\brr}{\begin{array}}
\newcommand{\err}{\end{array}}
\newcommand{\bc}{\begin{center}}
\newcommand{\ec}{\end{center}}

\def\A{{\cal A}}
\def\B{{\cal B}}

\def\D{{\cal D}}
\def\E{{\cal E}}
\def\G{{\cal G}}

\def\H{{\cal H}}
\def\N{{\cal N}}

\def\L{ \Lambda}
\def\N{{\cal N}}
\def\S{{\cal S}}

\def\P{{\cal P}}

\def\T{{\cal T}}

\def\F{{\cal F}}
\def\l{\lambda}
\def\s{\sigma}
\def\al{\alpha}
\def\be{\beta}
\def\ga{\gamma}
\def\de{\delta}
\def\ep{\epsilon}
\def\o{\omega}
\def\co{\Omega}
\def\th{\theta}
\def\var{\varphi}
\def\z{\zeta}
\newcommand{\bea}{\begin{eqnarray}}
\newcommand{\eea}{\end{eqnarray}}
\newcommand{\bean}{\begin{eqnarray*}}
\newcommand{\eean}{\end{eqnarray*}}
\newcommand{\CC}{\Bbb C}

\newcommand{\ZZ}{\Bbb Z}
\newcommand{\TT}{\Bbb T}

\newcommand{\del}{\partial}

\begin{document}

\begin{titlepage}

\begin{flushright}
CBPF--NF--061/97\\
\end{flushright}

\vskip0.5cm
\centerline{\LARGE Vertex Operator Representation of the Soliton}

\centerline{\LARGE Tau Functions in the $A_n^{(1)}$ Toda Models}

 \centerline{\LARGE by Dressing Transformations }

\vskip1.5cm
\centerline{\large   H. Belich
\footnote{E--mail address belich@cbpfsu1.cat.cbpf.br},
 G. Cuba
\footnote{E--mail address gcubac@cbpfsu1.cat.cbpf.br}} 
\vskip0.5cm
\centerline{\large and }
\vskip0.5cm
\centerline{\large R. Paunov
\footnote{E--mail address paunov@cbpfsu1.cat.cbpf.br}}
\centerline{Centro Brasileiro de Pesquisas Fisicas }
\centerline{Rua Dr. Xavier Sigaud 150, Rio de Janeiro, Brazil}

\vskip2.5cm
\abstract{ We study the relation between the group--algebraic approach and the 
dressing symmetry one to the soliton solutions of the $A_n^{(1)}$ 
Toda field theory in $1+1$ dimensions. Originally solitons in the affine
Toda models has been found by Olive, Turok and Underwood. Single 
solitons are created by exponentials of elements which ad--diagonalize
the principal Heisenberg subalgebra. Alternatively Babelon and
Bernard exploited the dressing symmetry to reproduce the known
expressions for the fundamental tau functions in the sine--Gordon model.
In this  paper we show the equivalence between these two 
methods to construct solitons in the $A_n^{(n)}$ Toda 
models.}

\end{titlepage}

\section{ Introduction}

\setcounter{equation}{0}
\setcounter{footnote}{0}

Exact localizable solutions of a relativistic 
field theory with finite energy are known in the 
literature as (depending on the dimension of the space--time)
solitons (kinks) or monopoles \cite{Raj}. Such solutions
play an important role,  since they are interpreted as new
particles, which appear in the spectrum of the theory.
As argued in \cite{ Raj, FaKo}, the particles corresponding
to solitons (or monopoles) are of nonperturbative nature. 
Usually, the solitons (monopoles) are also characterized 
by nontrivial topological charges. This provides a  
connection between the field theory and the geometry \cite{NS}. 
There exists \cite{Col} a remarkable duality which exchanges 
the sine--Gordon model with the massive Thirring model.
Under this duality transformation 
perturbative particles are mapped into solitons and 
vice versa.  The duality also  exchanges 
Noether currents with topological ones. Recently \cite{SW}, 
Seiberg and Witten has shown that there exist supersymmetric
field theories in four dimensions which exhibit exact duality.

 Within the Inverse Scattering Method (ISM) \cite{Ab}, soliton 
equations are required to admit Lax or zero curvature 
representation. This representation guarantees that 
the spectrum of the Lax operator is constant in time.
The leading idea of the ISM is to consider the time 
evolution as an evolution of the scattering data of the
corresponding Lax operator. In view of the zero curvature 
condition,  the equations of motion for 
the associated scattering data can be found explicitly.
In the framework of the ISM, solitons correspond 
to vanishing reflection coefficients. Imposing this 
condition, the inverse spectral transformation 
reduces to a linear algebraic system. The integrability 
(in the sense of Liouville) also ensures that the interaction
 between the single solitons is elastic. This intriguing 
property survives the quantization and was used to 
get exact quantum $S$ matrices in various integrable
 models \cite{Zam}.

The propagation of waves in one--dimensional lattices 
with exponential nearest--neighbour interaction was studied
 by Toda \cite{T}. In this pioneering paper, exact soliton 
solutions were also studied. The Toda lattice equations 
admit a field theoretical analogue in $1+1$ dimensions 
\cite{OT} which exhibit both the integrability and the relation
to the Lie algebras. Due to the last sequence of papers,
it became clear that (generalized) Cartan matrices can be 
used to construct integrable exponential interactions in 
two dimensions. It was  also shown that the 
field equations admit   zero curvature representation 
of a Lax connection whose components belong to a  certain 
Lie algebra. A deep relation between integrable 
hierarchies and the Kac--Moody (or affine) Lie algebras
\cite{Kac} has been clarified by Drinfeld and Sokolov
\cite{DS}. In the last paper it was also explained the {\it crucial} 
role of Heisenberg subalgebras and the related to them gradations
in constructing integrable evolution  equations \cite{gDS, Fer}.

Soliton solutions of the sine-Gordon 
model are known since the early days of the ISM \cite{Ab}.
$N$--soliton solutions in the $A_n^{(1)}$ Toda models
are found by the use of the Hirota method in \cite{Hol}.
It also became clear that physical observables, when 
evaluated on solitons, take finite real quantities for imaginary
values of the coupling constant. These results has been
extended by using group--algebraic methods \cite{OlS, Kne}, 
which also provide a bridge between the integrable models 
and the Kac--Moody algebras. In particular, the soliton
tau functions corresponding to a fundamental highest weight state
$|\L>$ of the affine Lie algebra admit the representation
\a
\frac{\tau_{\L}(\Phi)}{\tau_{\L}(\Phi_0)} = 
<\L|\prod_{i=1}^N(1+X_i F^{r_i}(\mu_i))|\L>
\label{1.1} 
\b             
where $\Phi_0$ is the vacuum solution, $X_i$ are numerical
 factors depending exponentially on the light cone 
coordinates and $F^{r_i}$ are elements of the affine 
Lie algebra which diagonalize the adjoint action of
the principal Heisenberg subalgebra (for details
see \cite{Kac, OlS} ) and $N$ is the number of the solitons. The above expression  is analogous 
to the representation obtained by the Kyoto group 
\cite{jap} for the tau function of the KP hierarchy.
 Integrable models possess a dressing symmetry 
\cite{STS}. The dressing group acts by gauge 
transformations on the Lax connection and preserves
 its form. An important property of the dressing 
symmetry is that it is a Lie--Poisson group. 
Applications of this symmetry to the Toda field
 theories has been done in \cite{fr}. Using the
 dressing symmetry, an alternative expression 
for the soliton tau functions arises
\a
\frac{\tau_{\L}(\Phi)}{\tau_{\L}(\Phi_0)} = 
<\L|g_-^{-1}( x^+, x^-).g_+( x^+, x^-)|\L>
\label{1.2}
\b
where $g_-^{-1}$ and $g_+$ are triangular elements of the affine group  which generate solitons from the vacuum. Babelon
 and Bernard  demonstrated that the expressions 
(\ref{1.1}) and (\ref{1.2}) are equivalent, at least 
for affine sine--Gordon solitons. In a
previous paper \cite{Bel} we obtained explicit 
expressions for the elements $g_\pm$ (\ref{1.2}). In 
the present work, by using the results of \cite{Bel},
we show that the vertex operator representation 
of the  tau functions (\ref{1.1}) corresponding to fundamental representations of the affine Lie algebra  is a consequence of the dressing group expression
 (\ref{1.2}) for arbitrary $A_n^{(1)}$ Toda solitons.

 The paper is organized as follows: In Sec. $2$ we 
introduce the $A_n^{(1)}$ Toda models, the soliton 
solutions and briefly comment the results of \cite{Bel}.
 Sec. $3$ is also complementary: the vertex operator
 representation of the $A_n^{(1)}$ algebras in the 
principal gradation is derived. In Sec. $4$ we obtain
 the vertex operator representation of the monosoliton
 tau functions starting from the dressing symmetry.
Sec.5 generalizes this result to generic soliton
 solutions. The Appendix contains a brief summary of the Lie
algebraic background used in the paper.

\section{ $A_n^{(1)}$ Toda Solitons and the Related Dressing Transformations}

\setcounter{equation}{0}
\setcounter{footnote}{0}

 Solitons in the $A_n^{(1)}$ Toda models has been found by Hollowood \cite{ Hol} who used the Hirota equations. Physically relevant solutions appear  for imaginary values of the coupling constant, nevertheless the components of the Toda field are complex. Affine Toda field theories and their soliton solutions are also studied from the point of view of the ISM \cite{ Nid}. The underlying Jost solutions \cite{Ab} as well as the elements of the transition matrix loose, in general, their nice analyticity properties. Similar phenomenon occurs when one considers the scattering data related to a linear differential operator of an arbitrary order \cite{Tomei} . In \cite{Bel} we developed an elegant  method to get $A_n^{(1)}$ Toda solitons. It strongly exploits the fact that the Lax connection belongs to the $A_n^{(1)}$ Lie algebra in the {\it principal} gradation (for a summary of the Lie algebraic background, see the Appendix).  The dynamical variables which we use to describe the soliton dynamics, appeared previously in the study of the periodic solution of the KdV equation and of the periodic Toda chain \cite{FMc}. The same approach, applied to the sine--Gordon solitons, was recently exploited to compute form factors in the quantum theory \cite{Bab}.

To get the $A_n^{(1)}$ Toda equations, we impose the zero
 curvature condition
\ai
\del_+A_--\del_-A_++\left[A_+, A_-\right]=0\label{2.1a}
\b
on the Lax connection
\a
A_\pm&=& \pm \del_\pm \Phi + m  e^{\pm ad\Phi}\E_{\pm}\0\\
 \del_{\pm}&=&\frac{\del}{\del x^\pm}
\label{2.1b}
\b
which belongs to the loop Lie algebra $\widetilde{\G}=\widetilde{SL}(n+1)$;
$x^{\pm}=x \pm t$ are the light cone coordinates in the two dimensional Minkowski space.
The elements $\E_\pm=\E_{\pm1}$ are generators of grade
$\pm 1$ of the principal Heisenberg subalgebra (\ref{A.7}), 
(\ref{A.23})
\a
\E_k=\l^k \sum_{p\in {\ZZ}_{n+1}} |p><p+k| \label{2.1c}
\b
and $\Phi$ is a diagonal matrix
\a
& &\Phi=\frac{1}{2}\sum_{i\in{\ZZ}_{n+1}}\var_i E^{ii}
\label{2.1d}
\bj
The $A_n^{(1)}$ Toda models has a conformally invariant
 extension \cite{cat} known as the $A_n$ Conformal Afine Toda 
(CAT) model. It also admits a zero curvature representation
 (\ref{2.1a}) for a connection of the form  (\ref{2.1b})
 in the {\it affine} Lie algebra $\hat{\G}=\hat{sl}(n+1)=A_n^{(1)}$
 \cite{Kac}. The affine algebra analogue of  (\ref{2.1d}) is
\a
\Phi \rightarrow \Phi+\eta \hat{d}
+\frac{\hat{c}}{2(n+1)}\zeta
\label{2.2}
\b
where (see the Appendix) $\hat{d}$ and $\hat{c}$ are 
the derivation and the central element respectively.
 The $A_n$ CAT field equations  are
\a
& &\del_+ \del_- \var_i =m^2 e^{2\eta}(e^{\var_i-\var_{i+1}}
-e^{\var_{i-1}-\var_i})\0\\
& &\del_+ \del_- \eta =0\0\\
& &\del_+ \del_- \zeta =m^2e^{2\eta}\sum_{i\in {{\ZZ}_{n+1}}}
e^{\var_i-\var_{i+1}}\0\\
\label{2.3}
\b
The first of the above equations coincides with 
the $A_n^{(1)}$ Toda equations provided that $\eta=0$.
As noted in \cite{BB}, affine Toda solitons arise 
after imposing the last restriction.
In such a case  (\ref{2.3}) admit a Hirota bilinear 
representation \cite{Hol}
\ai
& &\del_+\tau_k\del_-\tau_k-\tau_k\del_+\del_-\tau_k=
m^2(\tau_{k+1}\tau_{k-1}-\tau_k^2)\label{2.4a}\\
& &e^{-\var_k}=\frac{\tau_k}{\tau_{k-1}}\,\,\,\,,k\in{\ZZ}_{n+1}
\label{2.4b}\\
& &e^{\zeta_0-\zeta}=\prod_{k\in{\ZZ}_{n+1}}\tau_k
\label{2.4c}
\b
where
\a
\zeta_0={(n+1)}m^2x^+x^-
\label{2.4d}
\bj
The above value of the field $\zeta$ together with 
$\var_k=0, {k\in {\ZZ}_{n+1}}$ corresponds to the 
{\it vacuum} solution of the CAT model. The 
$A_n^{(1)}$ Toda vacuum is obtained by ignoring 
the additional field $\zeta$.
In \cite{Bel} we used an alternative procedure 
to get $A_n^{(1)}$ Toda solitons. The approach 
used by us strongly relies on the work \cite{Date} 
where several soliton equations, including the
sine--Gordon equation, has been studied. The dynamics of the
 $A_n^{(1)}$ Toda solitons is governed by 
the following {\it algebraic} equations
\ai
&&\prod_{l=1}^N \frac{\ep_{kl}+\o^{r_j}\mu_j}
{\ep_{kl}+\mu_j}=c_j \o^{r_j(1-k)}
\frac{e(\o^{r_j}\mu_j)}{e(\mu_j)}\0\\
&&e(\l)=exp \{m(\l x^++\frac{x^-}{\l})\}
\label{2.5a}\\
&&e^{-\var_k}= (-)^N \prod_{j=1}^N
\frac{\ep_{kj}}{\mu_j},~~~~k\in{\ZZ}_{n+1}
\label{2.5b}
\bj
where the integer $N$ stands for the number of solitons;
 $\mu_j, j=1, \ldots, N $ are (complex) parameters 
related to the soliton rapidities and $r_j$ are discrete 
parameters which take nonvanishing values in the cyclic
group ${\ZZ}_{n+1}$. The last are  known in the literature 
\cite{Hol, OlS, Kne} as {\it soliton species}. The relation between
solitons and $N$--body integrable relativistic
systems was studied in details \cite{OlA, body}. 
In order to simplify our analysis, only solitons 
with $|\mu_i| \neq |\mu_j|~(i \neq j)$ will be considered.
Certain particular solutions which violate this restriction
are also known \cite{BJ}.
Let us  note that in contrast to \cite{Hol, OlS, Kne},
we are working with a real value of the coupling constant.
This wants to say that the solutions (\ref{2.5a}), (\ref{2.5b})
are solitons in  {\it algebraic} sense.

Transformation groups for soliton equations were introduced in
 \cite{jap}. They has been also studied in relation to the underlying
 Riemann problem \cite{ZS} and are known as  groups of dressing transformations. In the present paper we will not comment the  
Poisson--Lie properties \cite{STS, fr} of the
dressing group.
For \dges\, which create monosoliton solutions from the vacuum in the sine--Gordon model this problem has been discussed in \cite{GR}.  
One believes  that \cite{BB}  in the limit 
$N \rightarrow \infty $, the $N$--soliton solutions are dense 
in the \dg\, orbit of the vacuum, and therefore, the
soliton parameters are expected to provide a convenient coordinate
system on the dressing group.

To introduce the group of dressing transformations
 we first recall that the zero curvature condition
 (\ref{2.1a})  can be equivalently written as
 the compatibility condition of the linear system
\a
(\del_{\pm} + A_{\pm})T=0
\label{2.6}
\b 
 We shall also impose the normalization condition 
$T|_{x^+=x^-=0} =1$. An element of the dressing group is represented by
 a pair of triangular elements $(g_+(x),g_-(x))$ of the 
corresponding loop (or affine) Lie group. This want to say
 that $g_\pm(x)=e^{\H}e^{\N_{\pm}}$ where $\H$ is the 
Cartan subalgebra and $\N_{\pm}$ contain all the elements
 of positive (negative) grade. The gradation is introduced by 
the derivation $\hat{d}$ of the loop (affine) Lie algebra (\ref{A.18}), (\ref{A.19}).
 The dressing group acts on the components of the Lax 
connection by gauge transformations
\ai
A_\mu \rightarrow A_\mu^g=-\del_\mu g_\pm g_\pm^{-1}+
g_\pm A_\mu g_\pm^{-1}
\label{2.7a}
\b   
 or equivalently 
\a
T(x) \rightarrow T(x)^g=g_{\pm}(x)T(x)g_{\pm}^{-1}(0)
\label{2.7b}
\bj
The space--time independent factors $g_{\pm}^{-1}(0)$ are added 
to ensure the normalization condition $T|_{x^+=x^-=0} =1$.
The gauge transformations are also
 required to preserve the form of the Lax connection
 (\ref{2.1b}). Therefore, the dressing transformations 
 form a symmetry group of the corresponding field equations
  (\ref{2.3}). Since the group elements $g_+$ and $g_-$ 
produce the same result  (\ref{2.7a}), it turns out that they
 are solution of the factorization problem
\a
g_-^{-1}(x)g_+(x)=T(x)g_-^{-1}(0)g_+(0)T^{-1}(x)
\label{2.8}
\b
 Comparing (\ref{2.1b}) with  (\ref{2.7a}), one observes that $g_+$
 and $g_-$ have opposite components on the Cartan torus $e^\H$. 
Therefore, the solution of  (\ref{2.8}) 
is unique. The multiplication in the dressing group 
 is the same as in the dual of the loop group\cite{STS, fr}. 
In particular, the map 
$(g_+,g_-)\rightarrow  g_-^{-1}g_+$ is not an isomorphism
 of  Lie groups.

In the CAT models based on an arbitrary simple Lie algebras,
one associates a group--algebraic tau function to each highest weight
vector $|\L>$ 
\a
\tau_{\L}(\Phi)=<\L|e^{-2\Phi} |\L>
\label{2.9}
\b
 Suppose that a solution $\Phi$ (\ref{2.2}), (\ref{2.3})
 with $\eta=0$ is related to the vacuum solution $\phi_0=\frac{m^2}{2}x^+x^-\hat{c}$
 by dressing transformation $(g_+,g_-)$. The following relation 
\cite{fr, BB} 
\a
\frac{\tau_{\L}(\Phi)}{\tau_{\L}(\phi_0)}=
<\L|g_-^{-1}(x)g_+(x) |\L>
\label{2.10}
\b
 between the tau functions of the two solutions is valid.
 In \cite{BB} it was observed that, after factorizing 
out the contribution which belongs to the center of the 
affine group, one can perform the calculation of 
the element $g_\pm$ in the corresponding loop group:
\a
g_{\pm}(x)=e^{\pm \frac{\zeta_0 - \zeta}{2(n+1)}\hat{c}}
\widetilde{g}_\pm(x)
\label{2.11}
\b
where $\widetilde{g}_\pm$, considered as elements of the
loop group, generate the $A_n^{(1)}$ Toda solution 
$\Phi=\frac{1}{2}\sum_{i\in{\ZZ}_{n+1}}\var_i |i><i|$
 from the vacuum $\Phi_0=0$. We recall that the above identity has been established in \cite{BB} for the 
affine Lie algebra $A_1^{(1)}=\hat{sl}(2)$. However, 
it is easy to check that the proof can be easily generalized for an arbitrary affine Lie algebra.

The representation (\ref{2.5a}), (\ref{2.5b}) of the $N$--soliton
 solutions was used  \cite{Bel} to calculate the elements 
 $\widetilde{g}_\pm$ which by (\ref{2.7a}) generate these solutions from the vacuum. It turns out that the dressing group elements  $\widetilde{g}_\pm$
 can be factorized into a product of monosoliton factors
\a
\widetilde{g}_\pm=\widetilde{g}_\pm(N)\widetilde{g}_\pm(N-1)
\ldots \widetilde{g}_\pm(1)
\label{2.12}
\b 
 where
\ai
\widetilde{g}_\pm(i)=e^{K(F_i)+P_i}e^{W_\pm(i)}
\label{2.13a}
\b
 In the above expression, $ F_i, K(F_i)$ and $P_i$ 
are diagonal traceless matrices
\a
& &F_i=\frac{1}{2}\sum_{k\in{\ZZ}_{n+1}} f_{ki}|k><k|~~~~
P_i=\frac{1}{2}\sum_{k\in{\ZZ}_{n+1}} p_{ki}|k><k| \0\\
& &K(F_i)=\sum_{k\in{\ZZ}_{n+1}}K_k(F_i) |k><k|\0\\
& &\sum_{k\in{\ZZ}_{n+1}} f_{ki}=\sum_{k\in{\ZZ}_{n+1}} p_{ki}=
\sum_{k\in{\ZZ}_{n+1}}K_k(F_i)=0
\label{2.13b}
\b
The quantities $K_k(F_i)$ obey the  relations
\a
K_{k}(F_i) - K_{k+1}(F_i)=\frac{f_{ki}+f_{ki+1}}{2}
\label{2.13c}
\b
which agree with the periodicity property $K_{k}(F_i)=K_{k+n+1}(F_i)$
since $F_i$ is traceless.
The loop algebra elements $W_\pm(i)$ are given by:
\a
& &W_{\pm}(i)=- K(F_i)+\sum_{k=1}^n f_{ki}\S_{\pm}^k(\mu_i)\0\\
& &\S_{\pm}^k(\mu_i)=\B_\pm^{k}(\mu_i) -\B_\pm^{n+1}(\mu_i)\0\\
& &\B_\pm^{k}(\mu_i)=\mp\left( \frac{1}{2}+\frac{(\frac{\l}{\mu_i})^{\pm(n+1)}}
{1-(\frac{\l}{\mu_i})^{\pm(n+1)}}\right) E^{kk} \mp \sum_{l\gtrless k}
\frac{(\frac{\l}{\mu_i})^{l-k\pm(n+1)}}{1-(\frac{\l}{\mu_i})^{\pm(n+1)}}E^{kl}\0\\
& &~~~~~~~~~~~~~~~~~~~\mp \sum_{l\gtrless k}\frac{(\frac{\l}{\mu_i})^{l-k}}{1-(\frac{\l}{\mu_i})^{\pm(n+1)}}E^{kl}
\label{2.13d}
\bj
In the above expressions $\l$ stands for the loop (or spectral) parameter
(see the Appendix).  Note that the elements $\B_{\pm}^{k}(\mu_i)$,
in contrast to $\S_{\pm}^k(\mu_i)$, are not in the loop algebra
$\widetilde{sl}(n+1)$. Due to the diagonal ($\l$--depending) contributions
 with nonvanishing trace, $\B_\pm^{k}(\mu_i)$ are in the loop algebra $\widetilde{gl}(n+1)$.
 Taking into account the loop algebra analogue of  (\ref{A.13}) together
 (\ref{A.23}) one gets
\a
\S_{\pm}^k(\mu_i)&=&\mp \sum_{r\in {\ZZ}_{n+1}} \o^r(\o^{-rk}-1)
\left(\frac{F_0^r}{2}+\sum_{p\gtrless 0}\frac{F_p^r}{\mu_i^p}\right) \label{2.14}
\b
 The elements of the Cartan subalgebra $K(F_i)$ (\ref{2.13b}),
 (\ref{2.13c}) can also be expanded in the alternative basis (\ref{A.13}).
 Therefore, from (\ref{2.13d}) and the above expansions, we obtain
\ai
& &W_{\pm}(i)= \sum_{k\in{\ZZ}_{n+1}} f_{ki}W_{\pm}^k(\mu_i)
\label{2.15a}
\b
 where 
\a
& &W_{+}^k(\mu_i)=-\sum_{r=1}^n\left(\frac{1-\o^{-rk}}{1-\o^{-r}}F_0^r
+\o^{r(1-k)}\sum_{p \geqslant 1}\frac{F_p^r}{\mu_i^p}\right)
\label{2.15b}\\
& &W_{-}^k(\mu_i)=\sum_{r=1}^n\left(-\frac{1-\o^{r(1-k)}}{1-\o^{-r}}F_0^r
+\o^{r(1-k)}\sum_{p \leqslant -1}\frac{F_p^r}{\mu_i^p}\right)
\label{2.15c}
\bj
 Note that neither $W_{+}^k(\mu_i)~$ nor $~W_{-}^k(\mu_i)$ contain
contributions belonging to the principal Heisenberg subalgebra.
Moreover, the identities
\a
W_{+}^k(\mu_i)-W_{-}^k(\mu_i)=-\sum_{r=1}^n\o^{r(1-k)}
\sum_{p \in {\ZZ}}\frac{F_p^r}{\mu_i^p}
\label{2.16}
\b
are valid. The above expression together with (\ref{2.15a})--(\ref{2.15c})
provides a hint of how to relate the dressing group approach 
to the group-algebraic methods, developed in \cite{ OlS, Kne}.
This relation has been conjectured for general 
 integrable hierarchies which admit a vacuum solution \cite{Fer}.

Up to now we discussed the  properties of the  
factors $\widetilde{g}_\pm(k)$ (\ref{2.12})  which produce single solitons without specifying
their relation to the $N$--soliton solution (\ref{2.5a}), (\ref{2.5b}).
This relation has been discussed in details in \cite{Bel}. Here we 
only list the results. First of all, it has been shown that the 
diagonal matrices $F_l,~l=1, \ldots ,N$ (\ref{2.13b}) satisfy
the identities
\ai
&&\sum_{k\in {\ZZ}_{n+1}} \o^{r(1-k)}
\left( \frac{\rho_{kl}(\o^r\mu_{l+1})}
{\rho_{kl}(\mu_{l+1})}-\o^{-r} \frac{\rho_{k+1l}
(\o^r\mu_{l+1})}
{\rho_{k+1l}(\mu_{l+1})}\right)\beta_k(F_{l+1})=\delta_{r, r_{l+1}}
(1-\o^r)\times\0\\
&&\times \prod_{a\neq l+1} 
\frac{\o^r\mu_{l+1}-\mu_a}
{\mu_{l+1}-\mu_a}
\prod_{a=1}^N \frac{ \mu_{l+1}+\ep_{1a}}
{\o^r\mu_{l+1}+\ep_{1a}}
\times\0\\
&& \times\sum_{k\in {\ZZ}_{n+1}}
\left( 1+\mu_{l+1} \frac{d}{d \l}
\ln \frac{\rho_{kl}}{\rho_{k+1l}}
(\mu_{l+1})\right)\beta_k(F_{l+1})
\label{2.17a}
\b
In the above system $r$ is a discrete parameter which 
belongs to ${\ZZ}_{n+1}$; $r_i=1, \ldots ,\, n$ stand for the 
soliton species (\ref{2.5a}), (\ref{2.5b}) related 
to the soliton with the rapiditylike parameter $\mu_i$ for $i=1, \ldots ,\,  N$. It was noted in \cite{Bel} that (\ref{2.17a}) is a system of algebraic equations which determines the variables
 \a
\beta_k(F_l)=e^{K_k(F_l)-\frac{f_{kl}}{2}}
\label{2.17b}
\b    
 The quantities $K_k(F_l)$ and $f_{kl}$ which appear in the above definition are introduced by (\ref{2.13a}) and (\ref{2.13c}). The functions  $\rho_{al}$ (\ref{2.17a}) on $\l$, $a\in {\ZZ}_{n+1},\,
l=1, \ldots ,\, N$ can be considered as components of $n+1$--dimensional
column vectors ${\bf \rho}_l(\l)$. The following  relations 
are valid
\a
{\bf \rho}_l(\l)&=&\D^{(l\,l-1)}(\l){\bf \rho}_{l-1}(\l)\0\\
{\bf \rho}_{l-1}(\l)&=&\D^{(l-1\,l)}(\l){\bf \rho}_{l}(\l)
\label{2.17c}
\b 
where $\D^{(l\,l-1)}(
\l)$ and $\D^{(l-1\,l)}(\l)$ are $(n+1)\times (n+1)$
matrices ($\D^{(l\,l-1)}(\l)\D^{(l-1\,l)}(\l)=1$). Their matrix elements are given by
\a
&&\D^{(l\,l-1)}_{ab}(\l)=\frac{\ga_l \be_b(F_l)}{(n+1)\rho_{bl-1}(\mu_l)}
\sum_{q\in {\ZZ}_{n+1}}\o^{q(a-b)} 
\frac{\l-\o^qe^{f_{bl}}\mu_l}{\l-\o^q\mu_l}\0\\
&&\D^{(l-1\,l)}_{ab}(\l)=\frac{\rho_{al-1}(\mu_l)}
{(n+1)\ga_l \be_b (F_l)}
\sum_{q\in {\ZZ}_{n+1}}\o^{q(a-b)} 
\frac{\l-\o^qe^{-f_{bl}}\mu_l}{\l-\o^q\mu_l}\0\\
&&\ga_l=\left( \prod_{p\in {\ZZ}_{n+1}}
\rho_{pl-1}(\mu_l))\right)^{\frac{1}{n+1}}\label{2.17d}
\b
The first relation (\ref{2.17c}) together with
\a
\rho_{j0}(\l)=\frac{1}{n+1}
\label{2.17e}
\b
determines uniquely the vectors ${\bf \rho}_l(\l)$.
The entries of the diagonal matrices $P_l$ are fixed by the equations
\cite{Bel}
\a
e^{K_a(F_{l-1})+\frac{p_{al-1}}{2}}&=&\frac{ \rho_{al-1}(\mu_l)}{\ga_l}
\label{2.17f}
\bj
 Therefore, the equations (\ref{2.17a})--(\ref{2.17f}) provide a 
recursive method to obtain the factorized expression (\ref{2.12})
 for a dressing group element in the loop group which generate the $N$--soliton 
solution (\ref{2.5a}), (\ref{2.5b}) from the vacuum. In view of the close relation (\ref{2.11})
between the dressing group elements in the affine group and in the
loop group, the decomposition  (\ref{2.12}) together with
the recurrence relations  (\ref{2.17a})-- (\ref{2.17f}), will
play a crucial role in obtaining the vertex operator 
representation for the soliton tau function. Note 
also that in view of (\ref{2.13c}), the quantities $\beta_k(F_i)$
(\ref{2.17b}) satisfy the relation 
\a
e^{-f_{kl}}=\frac{\beta_k(F_l)}{\beta_{k-1}(F_i)}
\label{2.18}
\b
which resembles (\ref{2.4b}). However $\beta_k$, as it can 
be seen from  (\ref{2.17a}) and (\ref{2.17c}), (\ref{2.17d}), do not satisfy 
the Hirota bilinear equations (\ref{2.4a}) in general. This wants to 
say that the fields $F_l$ (\ref{2.13b})  {\it do not}
satisfy the $A_n^{(1)}$ Toda equations for arbitrary values of $l$.         
  
The components $\rho_{al}(\l)$ of the vectors ${\bf \rho}_l(\l)$
(\ref{2.17a}), (\ref{2.17c}), (\ref{2.17d}) satisfy algebraic and 
differential equations which will be important in what follows.
First of all, combining (\ref{2.17c}) with (\ref{2.17d}), we get
\a
&&\sum_{a,r\in {\ZZ}_{n+1}}\frac{\o^{r(a'-a)}}{\l-\o^r\mu_l}
\left(\frac{\l-\o^re^{-f_{al}}\mu_l}{\l-\o^r\mu_l}\frac{\rho_{al}(\l)}
{\ga_l\be_a(F_l)}-
\frac{\rho_{al-1}(\l)}{\rho_{al-1}(\mu_l)}\right)=0
\label{2.19}
\b
for any $a'\in {\ZZ}_{n+1}$. From the above identity and the second 
equation (\ref{2.17d}) we deduce the relations
\a
&&\hskip -2cm \sum_{a\in {\ZZ}_{n+1}} \frac{d\, \D^{(l-1\, l)}_{a'a}}{d\l}(\l)
\rho_{al}(\l)=\frac{\rho_{a'l-1}(\mu_l)}{n+1}
\sum_{a,r\in {\ZZ}_{n+1}}\frac{\o^{r(a'-a)}}{\l-\o^r\mu_l}
\left( \frac{\rho_{al}(\l)}{\ga_l\be_a(F_l)}-\frac{\rho_{al-1}(\l)}
{\rho_{al-1}(\mu_l)}\right)\label{2.20}
\b
In view of (\ref{2.17c}), the following representation
\ai
&&{\bf \rho}_j(\l)=\D^{(j\, j-1)}\ldots \D^{(1\, 0)}{\bf \rho}_0(\l)
\label{2.21a}\b
takes place. Differentiating the above identity with respect to 
$\l$ and taking into account that ${\bf \rho}_0(\l)$ (\ref{2.17e})  does not depend on the spectral parameter, one obtains
\a
&&\frac{d\, {\bf \rho}_j}{d\l}(\l)=
\sum_{l=1}^j\D^{(j\, l)}(\l)\frac{d\, \D^{(l\, l-1)}}{d\l}(\l)    
{\bf \rho}_{l-1}(\l)\label{2.21b}
\b
where the matrices $\D^{(j\, l)}(\l)$ for $j\geq l$ are given by
\a
\D^{(j\, l)}(\l)&=&\D^{(j\, j-1)}(\l)\ldots \D^{(l+1\, l)}(\l)\0\\
{\bf \rho}_j(\l)&=& \D^{(j\, l)}(\l){\bf \rho}_l(\l)\0\\
\D^{(j\, j)}(\l)&=&1\label{2.21c}
\b
Since $\D^{(l-1\, l)}$ is the inverse of $\D^{(l\, l-1)}$, we can rewrite
(\ref{2.21b}) as follows
\a
&&\frac{d\, {\bf \rho}_j}{d\l}(\l)=
-\sum_{l=1}^j\D^{(j\, l-1)}(\l)\frac{d\, \D^{(l-1\, l)}}{d\l}(\l)    
{\bf \rho}_{l}(\l)\label{2.21d}
\bj
Inserting (\ref{2.20}) into the above identity we obtain
\a   
&&\hskip -2.5cm \frac{d\, \rho_{kj}}{d\l}(\l)=-\frac{1}{n+1}
\sum_{l=1}^j \sum_{a,a',r \in {\ZZ}_{n+1}}   
\frac{\o^{r(a'-a)}}{\l-\o^r\mu_l}\D^{(j\, l-1)}_{ka'}(\l)
\left( \frac{\rho_{al}(\l)}{\ga_l\be_a(F_l)}-\frac{\rho_{al-1}(\l)}
{\rho_{al-1}(\mu_l)}\right)\label{2.22}
\b
This result will be used in Sec. 5 to demonstrate the equivalence 
between the expressions (\ref{1.1}) and (\ref{1.2}) for the 
$N$--soliton tau functions.

\section{ Free field realization of the $A_n^{(1)}$ Lie algebras
in the Principal Gradation. }

\setcounter{equation}{0}
\setcounter{footnote}{0}

Realizations of infinite dimensional Lie algebras in terms of
harmonic oscillators play a crucial role in the representation
theory. They are also important in the applications to the
string theory and the two dimensional conformal models 
(for a review, see \cite{God}). In the simplest case of
the affine Lie algebra 
$A_1^{(1)}=\hat{sl}(2)$, the free field (or vertex operator)
construction was obtained in \cite{Lep}. This result has been
generalized for affine Lie algebras $\hat{\G}$ in the principal 
gradation \cite{Kaz} when the underlying classical 
Lie algebras $\G$ is simply laced and the level of the corresponding
representation is one. The generalization of the vertex operator
realization of $\hat{\G}$ with $\G$ being non--simply laced 
as well as for twisted affine Lie algebras, is also known 
\cite{OlS,Kne}. In the present Section we limit our attention 
on the $A_n^{(1)}$ Lie algebras in the principal gradation 
only. A rather detailed discussion of the general case is given
 in \cite{Kac}.

As it is seen from (\ref{2.10}), general solutions of the 
CAT models (\ref{2.1a}), (\ref{2.2}), (\ref{2.3})
can be written by using the highest weight representations
of the affine Lie algebras $\hat{\G}$. The representation spaces are generated by the action of 
arbitrary products of negative grade elements (\ref{A.19}) on 
the highest weight state $|\L>$. The last is annihilated by
 the elements of positive grade of the affine Lie algebra
\ai
& &F^r_k|\L>=0~~,\0\\
& &k \geqslant 1~~,~~~r \in {\ZZ}_{n+1}
\label{3.1a}
\b
 where we have used the notation (\ref{A.22}). We also recall
 that $F^0_{p(n+1)}\equiv 0~,~~p \in {\ZZ}$ since these elements
 are not in the affine algebra. The highest--weight state 
is  an eigenvector of the subalgebra $\hat{\G}_0$ of 
the elements of ${\ZZ}$ grade zero (\ref{A.18}), (\ref{A.19})
\a
F^r_0|\L>=\L(F^r_0)|\L>,~~~~\hat{d}|\L>=\L(\hat{d}),~~~~
\hat{c}|\L>=\L(\hat{c})|\L>\0\\
\label{3.1b}
\bj
 In view of (\ref{A.23}), the generators $\E_k=(n+1)F^0_k~~(k \neq 0 \, \rm{mod}(n+1))$ form a Heisenberg subalgebra. It is  known as the
{\it principal} Heisenberg subalgebra. Since $\E_k$ are bosonic
oscillators, they admit a Fock space representation, built up on the
Fock vacuum
\a
& &\E_k|0>=0,~~~~k \leqslant 1 \0\\
& &<0|\E_k=0,~~~~k \geqslant -1 \label{3.2}
\b 
It is known that when the value of the central charge
$\hat{c}$ is $1$, all the irreducible highest representations
of the Lie algebras $A_n^{(1)}$ are expressed in terms of
the elements of the (principal) Heisenberg subalgebra 
only \cite{Kac,Lep,Kaz}. In this Section we give an elementary demonstration of this statement.  

We start by introducing the notations
\ai
& &\E(\mu)=\sum_{\stackrel{k\in {\ZZ}}{k \neq 0 \, {\rm mod}(n+1)}}
\frac{\E_k}{\mu^k} \label{3.3a}\\
& &\th_r(\mu)=i\sum_{\stackrel{k\in {\ZZ}} {k \neq 0 \, \rm{mod}(n+1)}}
\frac{\o^{-rk}-1}{k}~~\frac{\E_k}{\mu^k}\0\\
& &\th_r(\mu)=\Phi_{r+n+1}(\mu)~,~~~\th_0(\mu)=0
\label{3.3b}
\b
where $i$  is the imaginary unit $i^2=-1$.
The above expressions are related by the identity 
\a
\E(\mu)=\frac{i}{n+1}\mu\frac{d}{d\mu}\sum_{r\in {\ZZ}_{n+1}}\th_r(\mu)
\label{3.3c}
\bj
 We also define the bosonic normal product $:\, :$ in the
standard way: $\E_k$ with $k \geqslant 1$ are moved
 to the right, while $\E_k$ with $\leqslant -1$ are moved
 to the left in each normal ordered monomial containing
 generators of the Heisenberg subalgebra only. Since the 
Heisenberg subalgebra does not contain elements of grade zero,
 it turns out that each normal ordered monomial of positive (negative) grade
 annihilates the Fock vacuum $|0>$ ($<0|$)
(\ref{3.2}). Taking into account (\ref{3.2}) and the Heisenberg
commutation relations (\ref{A.23}) we get the contraction
identities 
\ai
& &\th_r(\mu_1)\th_s(\mu_2)=<\th_r(\mu_1)\th_s(\mu_2)> + :\th_r(\mu_1)\th_s(\mu_2):\0\\
& &\E(\mu_1)\th_s(\mu_2)=<\E(\mu_1)\th_s(\mu_2)>+ :\E(\mu_1)\th_s(\mu_2):\0\\
&&|\mu_1|>|\mu_2|
\label{3.4a}
\b
The corresponding vacuum expectation
 values are given by 
\a
& &<\th_r(\mu_1)\th_s(\mu_2)>= \rm{ln} \frac{(\mu_1-\o^{-r}\mu_2)(\mu_1-\o^{s}\mu_2)}
{(\mu_1-\mu_2)(\mu_1-\o^{s-r}\mu_2)}\0\\
& &<\E(\mu_1)\th_r(\mu_2)>=i \frac{(1-\o^r\mu_2)\mu_1\mu_2}
{(\mu_1-\mu_2)(\mu_1-\o^{r}\mu_2)}
\label{3.4b}
\bj
 To obtain a realization of the basic representations of 
the affine Lie algebras $A_n^{(1)}$
we introduce the {\it vertex} operators
\a
V^r(\mu)&=&:\,e^{i\th_r(\mu)}\, :=e^{\sum_{k\leqslant -1}\frac{1-\o^{-kr}}{k}
~\frac{\E_k}{\mu^k}}e^{\sum_{k\geqslant 1}\frac{1-\o^{-kr}}{k}
~\frac{\E_k}{\mu^k}} \0\\
r&=&1,...,n
\label{3.5}
\b
Using the Wick theorem one deduces the 
following operator products:
\a
\E(\mu_1)V^r(\mu_2)&=&\frac{(\o^{r}-1)\mu_1\mu_2}
{(\mu_1-\mu_2)(\mu_1-\o^{r}\mu_2)}V^r(\mu_2)+:\E(\mu_1)V^r(\mu_2):\0\\
V^r(\mu_1)V^s(\mu_2)&=&\frac{(\mu_1-\mu_2)(\mu_1-\o^{s-r}\mu_2)}
{(\mu_1-\o^{-r}\mu_2)(\mu_1-\o^{s}\mu_2)}~
:e^{i(\th_r(\mu_1)+\th_s(\mu_2))}:\0\\
|\mu_1|&>&|\mu_2|\label{3.6}
\b
In view of (\ref{A.23}) and (\ref{3.2}) we also get
\a
\E(\mu_1)\E(\mu_2)&=&\frac{\mu_1\mu_2}{(\mu_1-\mu_2)^2}
-(n+1)\frac{\mu_1^{n+1}\mu_2^{n+1}}{(\mu_1^{n+1}-\mu_2^{n+1})^2}
+:\E(\mu_1)\E(\mu_2):\0\\
|\mu_1|&>&|\mu_2|\label{3.7}
\b
By using the notion of radial ordering \cite{God}
which is an analogue of the standard time ordering
in the field theory, one can extend the operator
products (\ref{3.6}) and (\ref{3.7}) to the region
$|\mu_1|<|\mu_2|$ in accordance to the "local" commutativity relations
\a
A(\mu_1) B(\mu_2)&=& B(\mu_2) A(\mu_1) \0\\
|\mu_1|&\neq &|\mu_2|\0
\b
for $A(\mu)$, $B(\mu)$ being $\E(\mu)$ (\ref{3.3b}) or $V^r(\mu)$ (\ref{3.5}).
This enables us to use the Laurent expansions 
\a
V^r(\mu)=\sum_{l\in {\ZZ}}\frac{V^r_l}{\mu^l}~~~,
~~~~~~V^r_l=\oint_{S^1}\frac{\rm{d}\mu}{2\pi i\mu}\mu^lV(\mu)
\label{3.8}
\b
Note also that normal ordered terms which appear in the 
right--hand side of (\ref{3.6}) and (\ref{3.7}) has finite matrix 
elements when evaluated between normalizable states 
in the bosonic Fock space.

To compute the algebra of commutators of the Laurent 
modes $\E_k,~~k\in {\ZZ}\backslash {(n+1)}{\ZZ}$ and $F^r_k$ we shall
use the {\it contour} deformation technique \cite{God}.
Let us briefly recall it: suppose that $A(\mu)$ and $B(\mu)$
are meromorphic operator valued functions on $\mu$ which
{\it locally} commute $A(\mu_1)B(\mu_2)=A(\mu_2)B(\mu_1)$ 
for $(|\mu_1|\neq|\mu_2|)$. Then the commutator between the
 Laurent coefficients
\ai
A_k=\oint_{S^1}\frac{\mu^k\rm{d}\mu}{2\pi i\mu}A(\mu)~~~,
~~~~~~B_l=\oint_{S^1}\frac{\mu^l\rm{d}\mu}{2\pi i\mu}B(\mu)
\label{3.9a}
\b
is given by
\a
\left[A_k,B_l\right]=\left(\oint_{|\mu_1|>|\mu_2|}-\oint_{|\mu_1|<|\mu_2|}
 \right)\mu_1^k \frac{\rm{d}\mu_1}{2\pi i\mu_1}~\mu_2^l
\frac{\rm{d}\mu_2}{2\pi i \mu_2}A(\mu_1)B(\mu_2)
\label{3.9b} 
\bj
Using the above identity, the Cauchy theorem and taking into account
(\ref{3.6}) one gets
\ai
& &\left[\E_k,V^r_l\right]=\left(\oint_{|\mu_1|>|\mu_2|}-\oint_{|\mu_1|<|\mu_2|}
 \right)\mu_1^k \frac{\rm{d}\mu_1}{2\pi i\mu_1}~\mu_2^l
\frac{\rm{d}\mu_2}{2\pi i \mu_2}\E_k(\mu_1)V^r(\mu_2)=\0\\
& &=\oint\mu_2^{l+1}\frac{\rm{d}\mu_2}{2\pi i \mu_2}
\left(\oint_{C_{\mu_2}}-\oint_{C_{\o^r\mu_{2}}}
 \right)\mu_1^{k+1} \frac{\rm{d}\mu_1}{2\pi i\mu_1}
(\o^r-1)\frac{V^r(\mu_2)}{(\mu_1-\mu_2)(\mu_1-\o^r\mu_2)}=\0\\
& &=(\o^{kr}-1)V^r_{k+l}
\label{3.10a}
\b
 where $C_\mu$ stands for a small anti--clockwise oriented contour 
which surrounds the point $\mu$. Similarly from the second 
equation (\ref{3.6}) we obtain the commutators
\a
& &\left[V^r_k,V^s_l\right]=\frac{(\o^r-1)(\o^s-1)}{(\o^{r+s}-1)}(\o^{sk}-\o^{rl})
V^{r+s}_{k+l}\0\\
& &r+s\neq 0~\rm{mod}(n+1)
\label{3.10b}\\
& &\left[V^r_k,V^{-r}_l\right]=-\o^r(1-\o^{-r})^2\left((\o^{-rk}-\o^{rl})\E_{k+l}
+k\o^{-rk}\delta_{k+l,0}\right)
\label{3.10c}
\bj   
 In deriving the above equations we have also used the identity
\a
\th_r(\o^p\mu)=\th_{p+r}(\mu)-\th_p(\mu)\0
\b   
which is a straightforward consequence from (\ref{3.3b}). Using the contour deformation technique one also concludes that (\ref{3.7}) is equivalent to the Heisenberg commutators (\ref{A.23}). Therefore we conclude that $\E_k,~~k\in {\ZZ}\backslash {(n+1)}{\ZZ}$ together with $V^r_k,\, r=1,...,n$,   $k\in {\ZZ}$ form a Lie algebra. The last is {\it isomorphic} to the Lie algebra $A_n^{(1)}$ in the principal gradation. To see this it is sufficient to set 
\a
F_k^r=\frac{\o^{rl}V_k^r}{(n+1)(\o^r-1)}\label{3.11}\\
l\in {\ZZ}_{n+1}, r=1,...,n\0
\b
 and to verify that (\ref{3.10a})--(\ref{3.10c}) together with (\ref{A.15}) coincide with (\ref{A.21}). In the above equation, $l\in {\ZZ}_{n+1}$ is an additional parameter. Note that the map
\a
\E_k &\longrightarrow  &\E_k\0\\
V_k^r &\longrightarrow  &\o^r V_k^r , r=1,...,n\label{3.12}
\b
is an automorphism of order $(n+1)$ of the Lie algebra (\ref{3.10a})--(\ref{3.10c}), (\ref{A.23}). It is clear that this automorphism is {\it outer}, and therefore the representations of $A_n^{(1)}$ corresponding to different values of the additional parameter $l$ (\ref{3.11}) are {\it inequivalent}. Due to (\ref{3.2}), (\ref{3.5}) and (\ref{3.11}) we obtain highest weight representations characterized by
\a
F_0^r |\Lambda_l>=\frac{\o^{rl}}{(n+1)(\o^r-1)}|\Lambda_l>\0\\
\widehat {c}=|\Lambda_l>=|\Lambda_l>,~~\widehat{d}|\Lambda_l>=0\label{3.13}
\b
where the highest--weight state coincides with the Fock vacuum (\ref{3.2}). Since $\widehat{c}=1$, we conclude that representations constructed by us are the $n+1$ {\it inequivalent fundamental} representations \cite{Kac} of the Lie algebra $A_n^{(1)}$ in the principal gradation.

We  proceed by calculating the tau functions (\ref{2.9}) in terms of the components of the field (\ref{2.1d}), (\ref{2.2}) with $\eta =0$. Taking also into account the last identity (\ref{A.13}), one obtains
\a
\Phi =\frac{1}{2}\sum_{\stackrel{1\leq r \leq n}{k\in{\ZZ}_{n+1}}}
\o^{r(1-k)}\var_k F_0^r + \frac{\zeta}{2(n+1)}\widehat{c}\label{3.14}
\b
Combining the above expression with (\ref{3.13}) we derive
\a
\left<\Lambda_l\left|\Phi\right|\Lambda_l\right>=\frac{\zeta}{2(n+1)}+\frac{1}{2(n+1)}\sum_{\stackrel{1\leq r \leq n}{k\in{\ZZ}_{n+1}}}\frac{\o^{r(l+1-k)}}{\o^r-1}\var_k\label{3.15}
\b
Therefore the group--algebraic tau functions (\ref{2.9}) satisfy the identities
\ai
\frac{\tau_{\L_k}(\Phi)}{\tau_{\L_{k-1}}(\Phi)}=e^{-\var_k},\,\,\,
k\in{\ZZ}_{n+1} \label{3.16a}\\
\prod_{k\in {\ZZ}_{n+1}}\tau_{\L_k}(\Phi)=e^{-\zeta}\label{3.16b}
\bj
Comparing the above identities with the definition (\ref{2.4a})--(\ref{2.4c}) of the Hirota tau funtions we get the relations
\a
\tau_{\L_k}(\Phi)=e^{-\frac{\zeta_0}{n+1}}\tau_k(\Phi)\label{3.17}
\b
where $\zeta_0$ is the vacuum value (\ref{2.4d}) of the field $\z$ (\ref{2.2}).

\section{ Derivation of the vertex operator representation for one--soliton tau
functions.}

\setcounter{equation}{0}
\setcounter{footnote}{0}

In the present Section we focus our attention on 
the equivalence between the representations (\ref{1.1})
and (\ref{1.2}) for one--soliton solutions
of the $A_n$ CAT equations. The existence of one--solitons,
i. e. solitary waves which propagate without changing 
their shape, is not a distinguishing property of the 
integrable evolution equations.
As a counter--example, one can quote the existence 
of kink solution of the $\var^4$ model in $1+1$ dimensions
\cite{Raj}. Monosoliton solutions of the affine Toda equations based on arbitrary simple Lie algebra are also known \cite{cat,Mac}.

In accordance with (\ref{2.5a}), (\ref{2.5b}) the 
 evolution of a single soliton is dictated by the equations
\a
& &e^{-\var_k}=\frac{1+X_k}{1+X_{k-1}}\0\\
& &X_k=-\frac{\o^{rk}}{c}~\frac{e(\mu)}{e(\o^r\mu)}
\label{4.1}
\b
where $\mu$ is the unique rapiditylike parameter,
$r= 1, \ldots , n$ is the soliton species and $c$
stands for the constant which appear in the 
right--hand side of (\ref{2.5a}). From the above
identities it follows that the corresponding
tau functions (\ref{2.4a}), (\ref{2.4b}) are given by
\a
\tau_k=1+X_k
\label{4.2}
\b 
 In this Section we shall demonstrate that the representations
 (\ref{1.1}) and (\ref{1.2}) for the one--soliton tau 
functions associated to the fundamental 
representations of the Lie algebra $A_n^{(1)}$ are equivalent.
Denote by $\widetilde{g}_{\pm}(\mu)$  the dressing group 
elements which generate the one--soliton solution (\ref{4.1})
of the $A_n^{(1)}$  Toda model  from the vacuum.
Taking into account  (\ref{2.12}) and (\ref{2.13a})
we get the expression
\a
& &\widetilde{g}(\mu)=\widetilde{g}^{-1}_{-}(\mu)\widetilde{g}_{+}(\mu)=
e^{-W_-{(\Phi)}}.e^{W_+(\Phi)}\0\\
& &\Phi=\frac{1}{2}\sum_{k\in {\ZZ}_{n+1}}\var_kE^{kk}
\label{4.3}
\b
where $\var_k$ are the components of the $A_n^{(1)}$  Toda
field and (\ref{2.15a}) $W_\pm(\Phi)=\sum_{k\in {\ZZ}_{n+1}}\var_kW_\pm^k(\mu)$.
 The elements $W_\pm(\mu)$ has been introduced by (\ref{2.15b}), 
(\ref{2.15c}). In what follows we shall use the representations
\ai
& &W_\pm(\Phi)=\sum_{k=1}^{n}\var_k \widetilde{W}_{\pm}^{(k)}(\mu)\0\\
& &\widetilde{W}_{\pm}^{(k)}(\mu)=W_{\pm}^{(k)}(\mu)-
W_{\pm}^{n+1}(\mu)
\label{4.4a}\\
& &\widetilde{W}_{\pm}^{(k)}(\mu)=-\sum_{k=1}^{n}\frac{1-\o^{-rk}}{1-\o^{-r}}
\oint_{\l \lessgtr |\mu|}\frac{\rm{d}\l}
{2\pi i\l}\frac{\o^{r}\l-\mu}{\l-\mu}F^r(\l)
\label{4.4b}
\b
 where 
\a
F^r(\l)=\sum_p\frac{F^r_p}{\l^p}~~~,
~~~~F^r_p=\oint\frac{\l^p\rm{d}\l}{2\pi i\l}F^r(\l)
\label{4.4c}
\bj
 To calculate the element (\ref{4.3}) in the fundamental 
 representation (\ref{3.11}), (\ref{3.13}) of the affine 
 Lie algebra $A_n^{(1)}$ we first introduce the notation
\ai
& &h(\al)=e^{-\al W_-(\Phi)}.e^{\al W_+(\Phi)}\0\\
& &h(\al)=\sum_{l=0}^{\infty}\frac{\al^l}{l!}
\frac{\rm{d}^l}{\rm{d}\al^l}h(0)
\label{4.5a}\\
& &\frac{\rm{d}^l}{\rm{d}\al^l}h(0)=
-W_-(\Phi)\frac{\rm{d}^{l-1}}{\rm{d}\al^{l-1}}h(0)+
\frac{\rm{d}^{l-1}}{\rm{d}\al^{l-1}}h(0).W_+(0)=\0\\
& &=\sum_{k=0}^{l}(-)^k\left( \begin{array}{cc}
l

\\

k  
\end{array}\right)       W_-^k(\Phi)W_+^{l-k}(\Phi)
\label{4.5b} 
\bj 
Using  (\ref{4.4a}), (\ref{4.4b}) and since the matrix (\ref{2.1d}) is traceless, one can also write
\a
W_{\pm}(\Phi)=\sum_{r=1}^{n}\frac{\widehat{\var}_r}{1-\o^{-r}}
\oint_{|\l|\lessgtr |\mu|}\frac{d\l}{2\pi i\l}\frac{\o^r\l-\mu}{\l-\mu}F^r(\l)\label{4.6}
\b
where we have introduced the discrete Fourier transform of $\var_k \,\, (\var_k=\var_{k+n+1})$ according to the expressions
\ai
\widehat{\var}_r=\sum_{k\in {\ZZ}_{n+1}}\o^{-kr}\var_k ;~~~\var_k=\frac{1}{n+1}\sum_{r\in {\ZZ}_{n+1}}\o^{kr}\var_r\label{4.7a}
\b
Let us also recall the identity
\a
\sum_{k\in {\ZZ}_{n+1}}\o^{k(i-j)}=(n+1)\delta_{i,j}^{(n+1)}\label{4.7b}
\bj
where the $\delta_{i,j}^{(n+1)}$ is the Kronecker symbol on the cyclic group: it vanishes for $i-j\neq 0 ~{\rm mod}\, (n+1)$ and $\delta_{i,j}^{(n+1)}=1$ if $i-j=0 ~{\rm mod}\, (n+1)$.

We proceed by calculating explicitly the Taylor expansion (\ref{4.5a}). First of all, we note that
\a
& &\frac{dh}{d\alpha}(0)=W_+(\Phi)-W_-(\Phi)=\0\\
& &=-\sum_{r=1}^{n}\frac{\widehat{\var}_r}{1-\o^{-r}}\oint_{C_{\mu}}\frac{d\l}{2\pi i\l}\frac{\o^r\l-\mu}{\l-\mu}F^r(\l)=\0\\
& &=-\sum_{r=1}^{n}\widehat{\var}_r\o^rF^r(\mu)\label{4.8}
\b
The above identities are in accordance with (\ref{2.15a}), (\ref{2.16}). To calculate the higher derivatives of $h(\alpha)$ (\ref{4.5a}) we first observe that due to (\ref{3.6}) one gets
\a
& &W_-(\Phi)V^r(\mu)-V^r(\mu)W_+(\Phi)=\0\\
& &=\frac{1}{n+1}\sum_{s=1}^{n}\frac{\o^{s(k+1)}\widehat{\var}_s}{(\o^{s}-1)^2}\left(\oint_{|\l|>|\mu|}-\oint_{|\l|<|\mu|}\right)\frac{d\l}{2{\pi}i\l}\frac{\o^s\l-\mu}{\l-\mu}V^s(\l)V^r(\mu)=\0\\
& &=\frac{1}{n+1}\sum_{s=1}^{n}\frac{\o^{s(k+2)}\widehat{\var}_s}{(\o^{s}-1)^2}\left(\oint_{|\l|>|\mu|}-\oint_{|\l|<|\mu|}\right)\frac{d\l}{2{\pi}i\l}\frac{\l-\o^{r-s}\mu}{\l-\o^r\mu}:e^{i(\theta_s(\l)+\theta_r(\mu))}:=\0\\
& &=\frac{1}{n+1}\sum_{s=1}^{n}\frac{\o^{s(k+1)}}{\o^{s}-1}\widehat{\var}_sV^{r+s}
(\mu),\,\,\,\,\,\, k\in {\ZZ}_{n+1}      \label{4.9}
\b
in the irreducible representation of $A_n^{(1)}$ with a highest--weight $|\Lambda_k>$ (\ref{3.11})--(\ref{3.13}). Using the above identity together with (\ref{4.5b}), it is easy to prove inductively that
\ai
& &\frac{d^lh}{d\alpha^l}(0)=(-)^l\sum_{s_1,...,s_l=1}^{n}\prod_{j=1}^{l}\Psi_{s_{j}}(\Phi).V^{s_1+...+s_l}(\mu)\label{4.10a}\\
& &\Psi_s(\Phi)=\frac{1}{n+1}\frac{\o^{s(k+1)}\widehat{\var}_s}{\o^{s}-1}\label{4.10b}
\bj
where we have omited the dependence on the parameter $k\in {\ZZ}_{n+1}$ which label the different fundamental representations of the affine Lie algebra. Substituiting the last result into the Taylor expansion (\ref{4.5a}), it turns out that (\ref{4.3}) is given by the following intermediate expression
\a
\widetilde{g}(\mu)=h(1)=\sum_{l=0}^{\infty}\frac{(-1)^l}{l!}\sum_{s_1,...,s_l=1}^{n}\prod_{j=1}^{l}\Psi_{s_{j}}(\Phi).V^{s_1+...+s_l}(\mu)\label{4.11}
\b
In order to write the right--hand side of the above identity in a more convenient form, we recall that in view of (\ref{3.3b}) and (\ref{3.5}), 
the vertex operators are periodic functions on the upper 
case index $V^{s}(\mu)=V^{s+n+1}(\mu)$ with $V^{0}(\mu)=1$.
 This observation suggests us to introduce the commutative 
associative algebra $\F $ of (complex) dimension $n+1$. It is 
generated by the elements  $V^{s}, ~s \in {\ZZ}_{n+1} $.
 The multiplication $\star$ in $ \F$ is defined by
\a
V^{r}\star V^{s}=V^{r+s}\label{4.12}
\b  
 Due to the periodicity condition, the element $V^{0}=V^{n+1}$ 
is the unity of $ \F$. It is worthwhile to note that (\ref{4.12})
 describe the {\it "fusion rules"} of a class of Rational Conformal
 Field Theoies \cite{Ver}. They correspond to a theory of a 
free massless scalar field in two dimensions which is compactified
 on circle with a rational value of the square of the radius. The 
algebra $\F$ can be endowed with a symmetric bilinear invariant
 form
\ai
<a,b>_{\F}&=&<b,a>_{\F}\0\\
<a,b\star c>_{\F}&=&<a\star b,c>_{\F}\0\\
a, b, c &\in& \F \label{4.13a}
\b
 which is uniquely fixed by
\a
<V^r,V^s>_{\F}=\delta^{(n+1)}_{r+s, 0}\label{4.13b}
\bj 
The algebra $\F$ together with the above bilinear 
form is an example of a Fr\"obenius algebra.
Fr\"obenius algebras and their deformations 
appear in the study of  the Topological
Field Theory \cite{Dub}. We go back to the problem
of simplifying the expression  (\ref{4.11}). Using
 the multiplication rules (\ref{4.12}) we can rewrite
 (\ref{4.11}) in the following form
\a
\widetilde{g}=\F~{\rm exp}\{-\sum_{s=1}^{n}\Psi_s(\Phi)V^s\}
\label{4.14}
\b  
where the symbol $\F$ means that the exponential is
taken in the Fr\"obenius algebra (\ref{4.12})--(\ref{4.13b}). In the above expression the dependence on $\mu$ 
of the elements $V^s,~s=1, \ldots, n$ was skipped. Note that
 the Fourier transformation  (\ref{4.7a}) diagonalizes 
the "fusion rules" (\ref{4.12})
\a
& &V^{r}\star \hat{V}^{s}=\o^{pr}\hat{V}^{p}\0\\
& &\hat{V}^{p} \star \hat{V}^{q}=(n+1)\delta^{(n+1)}_{p,q} \hat{V}^{p}
\label{4.15}
\b
 Substituing the above expression into (\ref{4.14}) 
we obtain
\ai
\widetilde{g}&=&\frac{1}{n+1}\sum_{p\in {\ZZ}_{n+1}}
e^{-\sum_{s=1}^n \o^{ps}\Psi_s(\Phi)}\hat{V}^{p}=\0\\
&=&\frac{1}{n+1}\sum_{p,r\in {\ZZ}_{n+1}}  \o^{-pr}.e^{\chi_p(\Phi)}V^r
\label{4.16a}\\
& &\chi_p(\Phi)=-\sum_{s=1}^n \o^{ps}\Psi_s(\Phi),\,\,\,
p\in {\ZZ}_{n+1}
\label{4.16b}
\bj
Due to the identity (\ref{4.7b}), the above quantities 
satisfy the restriction 
\ai
\sum_{p\in {\ZZ}_{n+1}}\chi_p(\Phi)=0
\label{4.17a}
\b 
Moreover, taking into account (\ref{4.10b}), we get 
the recurrence relations
\a
\chi_p(\Phi)-\chi_{p+1}(\Phi)=\var_{p+k+1}
\label{4.17b}
\bj
where $k\in {\ZZ}_{n+1}$ is the parameter which labels
the $n+1$ unequivalent fundamental representation of $A_n^{(1)}$
(\ref{3.11}), (\ref{3.13}). Comparing (\ref{4.17a}) and (\ref{4.17b}) 
with the definition (\ref{2.4a})--(\ref{2.4c}) of the Hirota tau
functions, we get 
\a
e^{\chi_p(\Phi)}=\frac{\tau_{p+k}(\Phi)}{(\prod_{l\in {\ZZ}_{n+1}}\tau_{l}(\Phi))^{\frac{1}{n+1}}}~~~~,~~~p\in {\ZZ}_{n+1}
\label{4.18}
\b 
Substituting back the above expression into (\ref{4.16a}) 
and taking into account (\ref{4.1}) as well as the vertex operator realization of the affine Lie algebra $A_n^{(1)}$ (\ref{3.5}),
(\ref{3.11}), (\ref{3.8}), we conclude that
\ai
& &\widetilde{g}(\mu)=\frac{1}{n+1}\sum_{p,r'\in {\ZZ}_{n+1}}\o^{-pr'}e^{\chi_p(\Phi)}V^{r'}(\mu)=\0\\
& &=\frac{1}{(n+1)\prod_{l\in {\ZZ}_{(n+1)}}\tau_{l}^{\frac{1}{n+1}}(\Phi)}\sum_{p,r'\in {\ZZ}_{n+1}}\left(\o^{-pr'}+\o^{p(r-r')+rk}X_0 F^r(\mu)\right)=\0\\
& &=\frac{1}{\prod_{l\in {\ZZ}_{(n+1)}}\tau_{l}^{\frac{1}{n+1}}(\Phi)}\left(1+(n+1)(\o^r-1)X_0 F^r(\mu)\right)\label{4.19a}
\b
From the above identity, (\ref{2.10}),  (\ref{2.11}) and  (\ref{2.4c})
one derives the relation
\a
& &\frac{\tau_{\Lambda_k}(\Phi)}{\tau_{\Lambda_k}(\Phi_0)}=e^{\frac{\zeta_0-\zeta}{n+1}}\left<\Lambda_k\left|{\widetilde{g}}_-(\mu){\widetilde{g}}_+(\mu)\right|\Lambda_k\right>=\0\\
& &\frac{e^{\frac{\zeta_0-\zeta}{n+1}}}{\prod_{l\in {\ZZ}_{(n+1)}}\tau_{l}^{\frac{1}{n+1}}(\Phi)}\left<\Lambda_k\left|\left(1+(n+1)(\o^r-1)X_0F^r(\mu)\right)\right|\Lambda_k\right>=\0\\
& &=\left<\Lambda_k\left|\left(1+(n+1)(\o^r-1)X_0F^r(\mu)\right)\right|\Lambda_k
\right>\label{4.19b}
\bj
Therefore, the representations (\ref{1.1}) and (\ref{1.2}) are equivalent in the one--soliton case.

\section{ $N$--soliton tau functions and vertex operators. }

\setcounter{equation}{0}
\setcounter{footnote}{0}

The existence of $N$-soliton solutions is a peculiar 
property of the integrable field theories.
From another point
of view \cite{body},  soliton solutions 
provide a relation of classes of solutions of integrable 
nonlinear partial differential equations with certain 
dynamical systems with a finite number of degrees of freedom. 
The last are known 
as $N$-body integrable systems (for a review, see \cite{OlA}). The
 $N$-soliton solutions, in the in-- and out-- limit, become 
asymptotically  a superposition of monosolitons separated 
in the space. Moreover, the transformation from the
 in-variables to the out-variables is canonical. The 
underlying generating function of this transformation
 is known as the classical S-matrix \cite{Ab}. An 
intriguing property of the classical S-matrix is 
that it is represented as a sum of terms, each of 
them representing a two-particle scattering. It is
 well known that the last property admits a 
generalization valid within the quantum theory:
  the quantum S-matrix which describes collision 
of solitons in a given quantum integrable 
model is a product of factors describing
 two particle interactions \cite{Zam}.
$N$--soliton solutions of the Toda field theories 
has been obtained both within the ISM \cite{Nid} and by using 
the Hirota method \cite{SP}

 We recall that the algebraic $N$-soliton
 solutions (\ref{2.5a}), (\ref{2.5b}) in the
 $A_n^{(1)}$ Toda Models are generated from the vacuum
by the dressing group elements (\ref{2.12})--(\ref{2.15c}).
 Due to the factorized expression (\ref{2.12}), we 
can also write
\a
\widetilde{g}&=&\widetilde{g}^{-1}_{-}\widetilde{g}_{+}
=\widetilde{g}^{-1}_{-}(1) 
\ldots \widetilde{g}^{-1}_{-}(N).\widetilde{g}_{+}(N) \ldots
\widetilde{g}_{+}(1)=\0\\
&=&\widetilde{g}(1)~.~{\rm Ad}~
\widetilde{g}^{-1}_{+}(1)
(\widetilde{g}(2)) \ldots {\rm Ad}(\widetilde{g}^{-1}_{+}(1))\ldots
\widetilde{g}^{-1}_{+}(N-1))
.(\widetilde{g}(N))\0\\
\widetilde{g}(i)&=&\widetilde{g}^{-1}_-(i) \widetilde{g}_+(i)
\label{5.1}
\b 
 In the above expression and in what follows 
we shall assume that the rapidity like parameters
 $\mu_i~~~,~~i=1, \ldots , N$ corresponding to the factors
  (\ref{2.13a})-(\ref{2.13d}) are 
radially ordered $|\mu_{1}|>|\mu_{2}|> \ldots >|\mu_{N}|$. 
In view of (\ref{4.16a}), (\ref{4.16b}) and the vertex
operator construction of the affine Lie algebra $A_n^{(1)}$
(\ref{3.5}), (\ref{3.8}), (\ref{3.11}), to calculate (\ref{5.1})
in the fundamental representations, one first has to obtain
an expression for the adjoint action of $\widetilde{g}_{+}(i)$
on the elements of the alternative basis (\ref{A.22}). To do that we first 
note that diagonal traceless $(n+1)\times(n+1)$ matrices
 $D=\sum_{k\in {\ZZ}_{n+1}}~d_k|k><k|$ are written in 
the alternative basis (\ref{A.13}), (\ref{A.15}) as 
follows
\a
D=\sum_{s\in {\ZZ}_{n+1}} \o^s \hat{d}_s F^s_0
\label{5.2}
\b
where $\hat{d}_s,~s\in {\ZZ}_{n+1}$  is the discrete
Fourier transformation (\ref{4.7a}) of the (diagonal) 
entries of $D$. Note that $F_0^0$ is {\it not} an element of
the Lie algebra $A_n$. However, since D is traceless, it
is clear that the coefficient $\hat{d}_0$ which multiplies $F_0^0$  vanishes. The above expression
 can be considered as an element either of the classical 
Lie algebra $A_n$ or of the affine algebra $A_n^{(1)}$. 
Taking into account the commutation relations  (\ref{A.21})
 together with (\ref{A.23}), it is easy to get
\ai
& &\left[F^r(\mu),D\right]=\frac{1}{n+1}
\sum_{k\in {\ZZ}_{n+1}}\frac{\o^s(\o^{sk}-1)\hat{d}_s}{\mu^k}F^{r+s}_k
\label{5.3a}   
\b 
 from where it is easy to get:
\a
&&e^{-D}F^r(\mu)e^{D}=\sum_{l=0}^{\infty} \frac{(-1)^l}{l!}
{\rm ad}^l D.F^r(\mu)=\0\\
&&=\sum_{k\in {\ZZ}}\frac{1}{\mu^k}\sum_{l=0}^{\infty}
 \frac{1}{l!}\sum_{s_1,\ldots ,s_l\in {\ZZ}_{n+1}}
\prod_{j=1}^l \o^{s_j} \frac{\o^{s_jk}-1}{n+1}
~\hat{d}_{s_j}~F_k^{r+s_1+\ldots +s_l}
\label{5.3b}
\bj
The right hand side of the above equation is of the same form as (\ref{4.11}).
This suggests to use the approach developed in Sec.4
to write (\ref{5.3b}) explicitly as a linear combination
of the affine algebra elements. In order to do that we first
introduce the following (reducible) representation of
the algebra $\F$ (\ref{4.12})
\a
V^s(F_k^r)&=&F^{r+s}_k\0\\
r,s &\in& {\ZZ}_{n+1}~~,~~~k\in {\ZZ}
\label{5.4}
\b
In complete analogy with the derivation of
the expression  (\ref{4.14}), we conclude that
\a
e^{-D}F^r(\mu)e^{D}=\sum_{k\in {\ZZ}_{n+1}}\frac{1}{\mu^k}
\F {\rm exp}(\sum_{s\in {\ZZ}_{n+1}}\frac{\o^s(\o^{sk}-1)}{n+1}\hat{d}_s V^s)
(F^r_k)\label{5.5}
\b
As  in Sec. 4, we diagonalize the operators $V^s$
(\ref{5.4}) 
\a
& &V^s(\hat{F}^p_k)=\o^{ps}\hat{F}^p_k\0\\
& &\hat{F}^p_k=\sum_r \o^{-pr}\hat{F}^r_k~~~~~~~~~~
F^r_k=\frac{1}{n+1}\sum_p\o^{pr}\hat{F}^p_k 
\label{5.6}
\b 
 Combining  (\ref{5.5}) with (\ref{5.6}) we obtain
\a
&&\hskip -1.5cm e^{-D}F^r(\mu)e^{D}=\frac{1}{n+1}\sum_{\stackrel{k\in {\ZZ}} {p \, \in {\ZZ}_{n+1}}}
\frac{\o^{pr}}{\mu^k}e^{d_{k+p+1}-d_{p+1}}\hat{F}^p_k=\0\\
&&\hskip -2cm = \frac{1}{n+1}\sum_{\stackrel{k\in {\ZZ}} {p,s \, \in {\ZZ}_{n+1}}}
\frac{\o^{p(r-s)}}{\mu^k} e^{d_{k+p+1}-d_{p+1}}\,F^s_k
=\frac{1}{n+1}\sum_{l,p,s \, \in {\ZZ}_{n+1}}
\o^{p(r-s)}\frac{e^{d_{l+p+1}}}{e^{d_{p+1}}}
\sum_{k\in {\ZZ}}\frac{F^s_{k(n+1)+l}}{\mu^{k(n+1)+l}}=\0\\
&&\hskip -1.5cm =\frac{1}{(n+1)^2}\sum_{l,p,q,s \, \in {\ZZ}_{n+1}} 
\o^{p(r-s)+lq}
\frac{e^{d_{l+p+1}}}{e^{d_{p+1}}}~F^s(\o^q\mu)
\label{5.7}
\b
As a consequence of the above identity and (\ref{2.17f}), 
 one obtains
\ai
e^{-{\rm ad}(K(F_i)+P_i)}F^r(\o^a\l)=
\sum_{s,q \in {\ZZ}_{n+1}}~U^{ra}_{sq}(i)F^s(\o^q\l)
\label{5.8a}
\b
where $K(F_i)$ and $P_i$ are diagonal traceless matrices (\ref{2.13b})
 and
\a
U^{ra}_{sq}(i)&=&M^{r+a}_{s+q}(i)N^a_q(i)\0\\
M^r_s(i)&=&\frac{\o^{s-r}}{n+1}\sum_{p\in {\ZZ}_{n+1}}
\frac{\o^{p(r-s)}}{\rho_{pi}(\mu_{i+1})}\0\\
N^a_q(i)&=&\frac{\o^{a-q}}{n+1} \sum_{p\in {\ZZ}_{n+1}}
\o^{p(q-a)} \rho_{pi}(\mu_{i+1})\label{5.8b}
\bj
In what follows we shall also need the commutators 
$\left[ W_+^k(\mu_i), F^s(\l)\right]$
where the affine algebra elements $ W_+^k(\mu_i)$ and $F^s(\l)$ 
are given by (\ref{2.15b}) and (\ref{4.4c}) respectively.
Using (\ref{A.21}) with $\hat{c}=1$ we get
\ai
\left[ W_+^k(\mu_i), F^s(\l)\right]&=&-\frac{1}{n+1}\sum_{r=1}^n
\frac{(1-\o^{-rk})\mu_i-\o^s(1-\o^{r(1-k)})\l}
{(1-\o^{-r})(\mu_i-\o^{s}\l)}~F^{r+s}(\l)+\0\\
&&+\frac{1}{n+1}\sum_{r=1}^n
\frac{(1-\o^{-rk})\mu_i-\o^{-r}(1-\o^{r(1-k)})\l}
{(1-\o^{-r})(\mu_i-\o^{-r}\l)}~F^{r+s}(\o^{-r}\l)-\0\\
&&-\frac{1}{(n+1)^2}\sum_{r=1}^n \frac{\o^{-rk}\l \mu_i}
{(\mu_i-\o^{-r}\l)^2}\de^{(n+1)}_{r+s,0},\,\,\, 
|\mu_i|>|\l|
\label{5.9a}
\b
 From the above expression, (\ref{2.13b}) and (\ref{2.15a})
 we also derive the commutator
\a
\left[ W_+(i), F^s(\l)\right]&=&\sum_{k \in {\ZZ}_{n+1}}
f_{ki}\left[ W_+^k(\mu_i), F^s(\l)\right]=\0\\
&&\hskip -3cm =\frac{1}{n+1}\sum_{r=1}^n\frac{\hat{f}_{ri}}{1-\o^{-r}}
\left( \frac{\mu_i-\o^{r+s}\l}{\mu_i-\o^{s}\l}F^{r+s}(\l)
-\frac{\mu_i-\l}{\mu_i-\o^{-r}\l}F^{r+s}(\o^{-r}\l)\right)-\0\\
&&\hskip -3cm -\frac{\hat{f}_{-si}}{(n+1)^2}\frac{\l \mu_i}
{(\mu_i-\o^{-s}\l)^2}~~~,~~~|\mu_i|>|\l|~,~~~s \in {\ZZ}_{n+1}
\label{5.9b}
\bj 
where $\hat{f}_{ri}$ stands for the discrete Fourier 
transformation (\ref{4.7a}) of $f_{ki}~,~~k\in {\ZZ}_{n+1}$.
Using (\ref{5.9b}) one proves inductively
\a
&&{\rm ad}^l W_+(i)F^s(\l)=\frac{\mu_i-\l}{\mu_i-\o^{s}\l}
\sum_{k=0}^{l}(-)^{l-k}\left( \begin{array}{cc}
l\\k  \end{array}\right) \times\0\\
&&\times \sum_{r_1, \ldots ,r_l=1}^n \frac{\mu_i-\o^{r_1+ \ldots +r_{k}+s}\l}
{\mu_i-\o^{-r_{k+1} - \ldots -r_{l}}\l}
\prod_{p=1}^l \psi_{r_pi}~~ F^{r_1+ \ldots +r_l+s}(\o^{-r_{k+1}- \ldots -r_{l}}\l)+\0\\
&&+\frac{\mu_i(\mu_i-\l)}{(n+1)^2(\mu_i-\o^{s}\l)}
\sum_{k=0}^{l-1}(-)^{l-k}\left( \begin{array}{cc}
l-1\\k \end{array}\right) \times\0\\
& &\sum_{r_1, \ldots ,r_{l-1}=1}^n \frac{\l\o^{-r_{k+1}- \ldots  -r_{l-1}}}
{(\mu_i-\o^{-r_{k+1}- \ldots -r_{l-1}}\l)
(\mu_i-\o^{r_{1}+ \ldots +r_k+s}\l)}~.~\hat{f}_{-r_{1}\ldots -r_{l-1}-s \, i}\0\\
&&\times \prod_{p=1}^l \psi_{r_p\, i}~~,~~~l \geqslant 1\0\\
&&~~~~~~~~~~~~~\psi_{r \, i}=\frac{\hat{f}_{r\, i}}{(n+1)(1-\o^{-r})}~~~,~~r=1,\ldots,n 
\label{5.10}
\b
Note that after the change $\hat{f}_{r\, i} \rightarrow \hat{\var}_{i}$, the quantities $\psi_{r\, i}$ 
coincide up to phase with  (\ref{4.10b}).
To write the right-hand side of the above identity
in a compact form, we consider the commutative associative 
algebra 
$\A=\F\times \T$. We recall that $\F$ is the Frobenius
algebra (\ref{4.12})--(\ref{4.13b})
while $\T$ is spanned on the elements $\TT^{r}~~,r \in {\ZZ}_{n+1}$.
The multiplication in $\A$ is introduced by
\ai
V^{r}* V^{s}&=&V^{r+s}~~~~,~~~\TT^{r} * \TT^{s}=\TT^{r+s}\0\\
V^{r}* \TT^{s}&=&\TT^{s}* V^{r} 
\label{5.11a}
\b 
We shall also need  specific representations of
 $\A$. They are defined by the relations
\a
V^{r}(g(\l)F^s(\l))&=&g(\l)V^{r}(F^s(\l))=g(\l)F^{r+s}(\l)\0\\
\TT^{r}(g(\l)F^s(\l))&=&g(\o^{-r}\l)\TT^{r}(F^s(\l))=g(\o^{-r}\l)
F^s(\o^{-r}\l)
\label{5.11b}
\bj
where $g$ is a function  on the complex parameter $\l$.
It is also assumed that the algebra $\A$ acts 
trivially on $\mu_i$  (\ref{5.10}). Using  (\ref{5.11a}) and  (\ref{5.11b}), it is easy to check that (\ref{5.10}) can be written as follows
\a
&&(-)^l{\rm ad}^l W_+(i)F^s(\l)=\0\\
&&\hskip -2cm =\frac{\mu_i-\l}{\mu_i-\o^{s}\l}
\left( (\sum_{r=1}^n\psi_{r\, i}V^{r}(\TT^{r}-1))^l_{*} \frac{\mu_iF^s(\l)}
{\mu_i-\l}-\o^{s}(\sum_{r=1}^n\o^r\psi_{r\, i}V^{r}(\TT^{r}-1))^l_{*}
\frac{\l F^s(\l)}{\mu_i-\l} \right)+\0\\
&&\hskip -2cm +\frac{\mu_i(\mu_i-\l)}{(n+1)^2(\mu_i-\o^{s}\l)}
\left<\sum_{p\in {\ZZ}_{n+1}}\hat{f}_{pi}V^{p},
(\sum_{r=1}^n\psi_{r\, i}V^{r}(1\otimes_{\T}\TT^{r}-
\TT^{r}\otimes_{\T}1))^{l-1}_{*}V^s\right>_{\F}\times\0\\
&&\times\frac{1}{\mu_i-\o^{s}\l}\otimes \frac{\l}{\mu_i-\l}
\label{5.12}
\b
where $\left<~,~\right>_{\F}$ is the invariant bilinear form  in the Fr\"obenius algebra $\F$ (\ref{4.12})--(\ref{4.13b}). We have also used the notation
\ai
\left(\P\right)_*^l=\underbrace{\P *...*\P}_l\label{5.13a}
\b
to indicate that the $l-$th power of $\P \in \T$ is taken with respect to the multiplication (\ref{5.11a}). In what follows we shall use the operators
\a
\T \exp{(\P)}=\sum_{l=0}^{\infty}\frac{\left(\P\right)_*^l}{l!}\0\\
\T \frac{\exp{\left(\P\right)-1}}{\P}=\sum_{l=1}^{\infty}\frac{\left(\P\right)_*^{l-1}}{l!}\label{5.13b}
\bj
Note that the second operator has eigenvalue one when applied on the zero modes of $\P$. Taking into account (\ref{5.12}) together with (\ref{5.13a}), (\ref{5.13b}) we obtain
\a
~~~\0\\
&&e^{-{\rm ad} W_+(i)}.F^s(\l)=\0\\
&&\hskip -2cm =\frac{\mu_i-\l}{\mu_i-\o^s\l}\left(\T e^{(\sum_{r=1}^n\psi_{ri}V^r({\TT}^r-1))}\frac{\mu_iF^s(\l)}{\mu_i-\l}
-\o^s\T e^{(\sum_{r=1}^n\o^r\psi_{ri}V^r({\TT}^r-1))}\frac{\l F^s(\l)}
{\mu_i-\l}\right)+\0\\
&&\hskip -2cm \frac{\mu_i(\mu_i-\l)}{(n+1)^2(\mu_i-\o^s\l)}
\left<\sum_{p\in {\ZZ}_{n+1}}\hat{f_{p_i}}V^p,
\T \frac{\exp{(\sum_{r=1}^n\psi_{ri}V^r(1{\otimes}_{\T}{\TT}^r-
{\TT}^{-r}\otimes_{\T} 1))}-1}
{\sum_{r=1}^n\psi_{ri}V^r(1\otimes_{\T}{\TT}^r-{\TT}^{-r}\otimes_{\T} 1)}V^s\right>_{\F}\times\0\\
&&\times\frac{1}{\mu_i-\o^s\l}\otimes\frac{\l}{\mu_i-\l}\label{5.14}
\b
In view of the above equation, we shall need the (reducible) representations (\ref{5.11b}) of $\T$ (\ref{5.11a})  spanned on the vectors $g(\o^r\l)F^s(\o^r\l),~ r,s\in {\ZZ}_{n+1}$ for $g$ being an arbitrary meromorphic function on $\l$. It is easy to check that
\ai
G^{pq}(\l)=\sum_{r,s\in {\ZZ}_{n+1}}\o^{-ps+rq}g(\o^r\l)F^s(\o^r\l)
\label{5.15a}
\b
are eigenvectors of $\T$
\a
V^k G^{pq}(\l)=\o^{pk}G^{pq}(\l),~~
{\TT}^k G^{pq}(\l)=\o^{kq}G^{pq}(\l).\label{5.15b}
\b
Due to the identity (\ref{4.7b}), the inverse of (\ref{5.15a}) is given by
\a
g(\o^r\l)F^s(\o^r\l)=\frac{1}{(n+1)^2}\sum_{p,q\in {\ZZ}_{n+1}}\o^{ps-rq}G^{pq}(\l)\label{5.15c}
\bj
Note also that
\ai
\sum_{r=1}^n\o^{rp}\psi_{r\, i}=\frac{f_{p\, i}}{2}-K_p(F_i)\label{5.16a}
\b
where the diagonal traceless matrices $F_i$ and $K(F_i)$ has been defined by (\ref{2.13b}). To obtain the above identity we have also used the  relations (\ref{2.13c}). Taking into account  (\ref{2.17b}) we can  write
\a
e^{-\sum_{r=1}^n\o^{rp}\psi_{r\, i}}=\beta_p(F_i)\label{5.16b}
\bj
substituing back (\ref{5.15a})--(\ref{5.16b}) into (\ref{5.14}) one gets
\a
& &e^{-ad W_+(i)}.F^s(\l)=\frac{\mu_i-\l}{(n+1)^2(\mu_i-\o^s\l)}\times\0\\
& &\times\sum_{p,q,r',s'\in {\ZZ}_{n+1}}\o^{p(s-s')+r'q}.\frac{\mu_i-\o^{r'+s'}\l}{\mu_i-\o^{r'}\l}
\frac{\be_p(F_i)}{\be_{p+q}(F_i)}F^{s'}(\o^{r'}\l)+\0\\
&&+\frac{\l \mu_i(\mu_i-\l)}{(n+1)^4(\mu_i-\o^s\l)}
\sum_{a,q,q'p,p'\in {\ZZ}_{n+1}}f_{ai}\frac{\o^{as+pq+p'(q'+1)}}
{(\mu_i-\o^{p+s}\l)(\mu_i-\o^{p'}\l)}\times\0\\
& &\times\frac{\be_{a-q}(F_i)-\be_{a+q'}(F_i)}{\be_{a+q'}(F_i)({\rm ln}\be_{a-q}(F_i)
-{\rm ln})\be_{a+q'}(F_i)}
\label{5.17}
\b
Therefore, the adgoint action of the affine group element $e^{- W_+(i)}$
on the affine  algebra, can be written as follows
\ai
&&e^{-ad W_+(i)}\,F^s(\o^c \l)=\sum_{q,v\in {\ZZ}_{n+1}}
W^{sc}_{qv}[F_i](i;\l)\,F^q(\o^v\l)+Z^{sc}[F_i](i,\l)\0\\
&& {\rm for}~~~~~~~~~~~~~|\mu_i|>|\l|
\label{5.18a}
\b
Due to (\ref{2.15a}), $W_+(i)$ are linear in the entries $f_{ki}$ of the 
diagonal matrices $F_i$  (\ref{2.13b}). This wants to say that the relations
\a
\sum_{q,v\in {\ZZ}_{n+1}}W^{sc}_{qv}[F_i](i;\l)\,W^{qv}_{kl}[-F_i](i;\l)&=&
\de^{(n+1)}_{s,k}\de^{(n+1)}_{c,l}\0\\
\sum_{q,v\in {\ZZ}_{n+1}}W^{sc}_{qv}[F_i](i;\l)\,Z^{qv}[-F_i](i;\l)+
Z^{sc}[F_i](i;\l)&=&0
\label{5.18b}
\b
are valid. Comparing (\ref{5.17}) with (\ref{5.18a}) we obtain
\a
W^{sc}_{qv}[F_i](i;\l)&=&K^{s+c}_{q+v}[F_i](i;\l)L^c_v[F_i](i;\l)\0\\
K^{s}_{q}[F_i](i;\l)&=&\frac{1}{n+1}
\frac{\mu_i-\o^q \l}{\mu_i-\o^s \l}
\sum_{p\in {\ZZ}_{n+1}}\o^{p(s-q)}\be_p(F_i)\0\\
L^c_v[F_i](i;\l)&=&\frac{1}{n+1}
\frac{\mu_i-\o^c \l}{\mu_i-\o^v \l}
\sum_{p\in {\ZZ}_{n+1}}\o^{p(v-c)}\be_p(-F_i)
\label{5.18c}
\b
and 
\a
Z^{sc}[F_i](i,\l)&=&\frac{\o^{c}\l \mu_i(\mu_i-\o^{c}\l)}
{(n+1)^4(\mu_i-\o^{s+c}\l)}
\sum_{\stackrel{a,b,p}{v,v'}} f_{pi}\frac{\o^{sa+(p-a)v+(b+1-p)v'+c(a-b-1)}}
{(\mu_i-\o^{v}\l)(\mu_i-\o^{v'}\l)}\times\0\\
&\times & \frac{\be_{a}(F_i)-\be_{b}(F_i)}{\be_{b}(F_i)({\rm ln}\be_{a}(F_i)-
{\rm ln}\be_{b}(F_i))}\label{5.18d}
\bj
Taking into account that the matrices $F_i$ (\ref{2.13b}) are traceless and the identity
\a
\sum_{\stackrel {p \, \in {\ZZ}_{n+1}} {1\leq r \leq n}}
\frac{\o^{(p-a)r}}{1-\o^r}f_{pi}&=&
-(n+1)\ln \be_a (F_i)\0
\b
we conclude that (\ref{5.18d}) can be alternatively written in the form
\ai
Z^{sc}[F_i](i,\l)&=&\frac{\l (\mu_i -\o^c \l)}{(n+1)^3(\mu_i -\o^{s+c} \l)}
\sum_{a,b,v \in {\ZZ}_{n+1}} 
\frac{\o^{a(s+c+v)-b(v+c)}}{\l-\o^v\mu_i} \frac{\be_a(F_i)}{\be_b(F_i)}+\0\\
&+&\frac{\l \o^c \de^{(n+1)}_{s,0}}{(n+1)(\mu_i-\o^{s+c}\l)}
\label{5.19a}
\b
After certain trivial algebraic manipulations involving the first term of the above expression and using (\ref{2.18}), 
(\ref{2.17d}) we arrive at the result
\a
Z^{sc}[-F_i](i,\l)&=&\frac{\l}{\ga_i (n+1)^2 (\o^{s+c}-\mu_i)}
\sum_{a,b \in {\ZZ}_{n+1}} \o^{(s+c)a-bc} \D^{(i\, i-1)}_{ab}(\l)
\frac{\rho_{b i-1}(\mu_i)}{\be_a(F_i)}+\0\\
&+&\frac{\l \o^c \de^{(n+1)}_{s,0}}{(n+1)(\mu_i-\o^{s+c}\l)}
\label{5.19b}
\bj
where $\ga_i$  are the constants defined by (\ref{2.17d}).

The equations (\ref{5.8a}), (\ref{5.8b}) and (\ref{5.18a}) provide an 
expression for the adjoint action of the element (\ref{2.13a}) on the 
affine Lie algebra
\ai
\widetilde{g}^{-1}_+(i)F^s(\o^c\l) \widetilde{g}_+(i)&=&
\sum_{r,v \in {\ZZ}_{n+1}} \left(R^{sc}_{rv}(i;\l)F^r(\o^v\l)+
U^{sc}_{rv}(i)Z^{rv}[F_i](i;\l)\right)=\0\\
&=&\sum_{r,v \in {\ZZ}_{n+1}} R^{sc}_{rv}(i;\l)\left(
F^r(\o^v\l)-Z^{rv}[-F_i](i;\l)\right)\0\\
|\mu_i|& > & |\l| \label{5.20a}
\b
where $R^{sc}_{rv}(i;\l)$ are the elements of the matrix 
$R(j;\mu_i)=U(j)W(j;\mu_i)$ which acts on tensor product
${\CC}^{n+1}\otimes {\CC}^{n+1}$
\a
&&R^{sc}_{rv}(i;\l)=\sum_{p,q \in {\ZZ}_{n+1}}
U^{sc}_{pq}(i)\, W^{pq}_{rv}(i;\l)=P^{c}_{v}(i;\l)\,
Q^{s+c}_{r+v}(i;\l)\0\\
&&P^c_v(i;\l)=\frac{1}{n+1}\sum_{p \in {\ZZ}_{n+1}}
\o^{(p-1)(v-c)} \frac{\l -\o^{-v}e^{-f_{pi}}\mu_i}{\l-\o^{-v}\mu_i}
\frac{\rho_{pi}(\mu_{i+1})}{\be_p(F_i)}\0\\
&&Q^s_r(i;\l)=\frac{\mu_i-\o^r \l}{(n+1)^2}
\sum_{a,b \in {\ZZ}_{n+1}} \o^{(a-1)s-br}
\frac{\be_b(F_i)}{\rho_{ai}(\mu_{i+1})}
\sum_{p \in {\ZZ}_{n+1}}
\frac{\o^{(b+1-a)p}}{\mu_i-\o^p\l}
\label{5.20b}
\bj
Note that the second identity   (\ref{5.20a}) is a consequence of
(\ref{5.18b}). Comparing the above equations with (\ref{2.17d})
and taking into account (\ref{2.18}) we obtain the following alternative expressions
\a
&&\hskip -1.5cm P^c_v(i;\l)=\frac{\ga_i}{n+1}\sum_{a,b \in {\ZZ}_{n+1}}
\o^{(a-1)v-(b-1)c}
\frac{1}{\rho_{a i-1}(\mu_i)} \D^{(i-1\, i)}_{ab} (\l)\rho_{bi}(\mu_{i+1})\0\\
&&\hskip -1.5cm Q^s_r(i;\l)=\frac{1}{(n+1) \ga_i}
\sum_{a,b \in {\ZZ}_{n+1}} \o^{(a-1)s-(b-1)r}
\frac{1}{\rho_{a i}(\mu_{i+1})} \D^{(i\, i-1)}_{ab} (\l)\rho_{bi-1}(\mu_{i})
\label{5.21}
\b 
which together with (\ref{4.7b}), (\ref{2.21c}) yield
\ai
&&\left( P(j;\l)\ldots P(k;\l)\right)^c_v=
\frac{\ga_j \ldots \ga_k}{n+1} \times\0\\
&&\times \sum_{a,b \in {\ZZ}_{n+1}} \o^{(a-1)v-(b-1)c}
\frac{1}{\rho_{a k-1}(\mu_k)} \D^{(k-1\, j)}_{ab} (\l)\rho_{bj}(\mu_{j+1}) \0\\
&&\left( Q(j;\l)\ldots Q(k;\l)\right)^s_r=
\frac{1}{(n+1)\ga_j \ldots \ga_k}\times \0\\
&&\times \sum_{a,b \in {\ZZ}_{n+1}} \o^{(a-1)s-(b-1)r}
\frac{1}{\rho_{a j}(\mu_{j+1})} \D^{(j\, k-1)}_{ab} (\l)\rho_{bk-1}(\mu_k)\0\\
&&{\rm for}~~~~ k\leq j \label{5.22a}
\b 
Therefore, as a consequence of (\ref{5.20b}), (\ref{2.21c}), (\ref{5.18c}) and
the above identities we get 
\a
&&\hskip -2.5cm \left( R(j;\mu_{j+1})\ldots R(k;\mu_{j+1})\right)^{p0}_{rv}=
\frac{1}{(n+1)^2}\sum_{a,a',b' \in {\ZZ}_{n+1}} 
\o^{(a-1)v+(a'-1)p-(b'-1)(r+v)}\times\0\\
&& \hskip 2.5cm \times \frac{\rho_{ak-1}(\mu_{j+1})}{\rho_{ak-1}(\mu_k)}
\D^{(j\, k-1)}_{a'b'}(\mu_{j+1})\frac{\rho_{b'k-1}(\mu_k)}{\rho_{a'j}(\mu_{j+1})}
\label{5.22b}
\bj
On the other hand, using (\ref{2.17d}) and (\ref{2.21c}), it is not difficult
to show that
\a
\D^{(j\, k)}_{ab}(\o^r \l)&=& \o^{r(a-b)} \D^{(j\, k)}_{ab}( \l) \0
\b
Inserting this identity into (\ref{5.22b}) and taking into account (\ref{2.17e}) one obtains
\a
&& \left(R(j;\mu_{j+1})\ldots R(1; \mu_{j+1})\right)^{p0}_{rv}=
\frac{\de^{(n+1)}_{v,0}}{n+1}
\sum_{a \in {\ZZ}_{n+1}} \o^{(a-1)(p-r)} 
\frac{\rho_{aj}(\o^r\mu_{j+1})}{\rho_{aj}(\mu_{j+1})} \0
\b
and therefore
\ai
&&\hskip -2cm \sum_{p\in {\ZZ_{n+1}}} \o^{p(1-k)} \left(R(j;\mu_{j+1})\ldots R(1; \mu_{j+1})\right)^{p0}_{rv}=
\o^{r(1-k)} \de^{(n+1)}_{v,0} \frac{\rho_{kj}(\o^r\mu_{j+1})}{\rho_{kj}(\mu_{j+1})}
\label{5.23a}
\b
Using similar arguments and (\ref{2.17b})--(\ref{2.22}) we also prove the 
identity
\a
&&\hskip -2.5cm \sum_{\stackrel{1\leq l \leq j}{p\in {\ZZ_{n+1}}}}  
\o^{p(1-k)} \left(R(j;\mu_{j+1})\ldots R(l; \mu_{j+1})\,
Z[-F_l](l;\mu_{j+1})\right)^{p0}=
-\frac{\mu_{j+1}}{n+1}\, \frac{ d\ln \rho_{kj}}{d \l} (\mu_{j+1})
\label{5.23b}
\bj
where the functions $Z^{sc}[\pm F_i]$ has been defined by (\ref{5.18a}), 
(\ref{5.18b}). 

Now we are ready to calculate  the product (\ref{5.1}). Inserting 
back (\ref{5.18a}) and (\ref{5.18a}) into (\ref{5.1}) we get
\a
&& {\rm Ad }\left( \widetilde{g}^{-1}_{+}(1) \ldots 
\widetilde{g}^{-1}_{+}(i-1)\right) F^p(\mu_i)=\sum_{r,v \in {\ZZ}_{n+1}} 
\left( R(i-1;\mu_i)\ldots R(1;\mu_i)\right)^{p0}_{rv} F^r(\o^v\mu_i)-\0\\
&&-\sum_{j=1}^{i-1} \left( R(i-1;\mu_i)\ldots R(j;\mu_i)\, Z[-F_j](j;\mu_i)\right)^{p0}\label{5.24}
\b
On the other hand one can repeat the procedure developed in Sec. 4 and to 
obtain a result analogous to (\ref{4.16a}), (\ref{4.16b})
\a
&& \hskip -2cm \widetilde{g}(i)=\widetilde{g}^{-1}_-(i)\, \widetilde{g}_+(i)=
\frac{\hat{\be}_0(F_i)}{n+1}+\sum_{p\in {\ZZ_{n+1}}}
(\o^p-1)\hat{\be}_p(F_i) F^p(\mu_i) 
\label{5.25}
\b
where $\hat{\be}_k(F_i)$, $k\in {\ZZ}_{n+1}$ is the discrete Fourier 
transformation (\ref{4.7a}) of $\be_k(F_i)$ (\ref{2.17b}). Inserting back the 
above expression into (\ref{5.24}) and taking into account (\ref{5.23a}),
(\ref{5.23b}) and (\ref{2.17a}) we conclude that
\ai
&& {\rm Ad }\left( \widetilde{g}^{-1}_{+}(1) \ldots 
\widetilde{g}^{-1}_{+}(i-1)\right)\, \widetilde{g}(i)=
Y_i \left(1+ X_i F^{r_i}(\mu_i)\right)\0\\
&&X_i=(n+1)(1-\o^{r_i})  
\prod_{a\neq i} 
\frac{\o^{r_i}\mu_i-\mu_a}
{\mu_{l+1}-\mu_a}
\prod_{a=1}^N \frac{ \mu_i+\ep_{1a}}
{\o^{r_i}\mu_i+\ep_{1a}}\0\\
&&Y_i=\frac{1}{n+1}\sum_{k\in {\ZZ}_{n+1}}
\left( 1+\mu_i \frac{d}{d \l}
\ln \frac{\rho_{ki-1}}{\rho_{k+1i-1}}
(\mu_i)\right)\beta_k(F_i)
\label{5.26a}
\b
Note that due to  (\ref{2.5a}), the quantities 
$X_i$ depend exponentially on the light cone variables $x^+$ and $x^-$.
Combining (\ref{5.26a}), (\ref{2.10}) and (\ref{2.11}) we deduce the relation
\a
&& \hskip -2cm \frac{\tau_{\L}(\Phi)}{\tau_{\L}(\Phi_0)}=
e^{\frac{\z_0-\z}{n+1}}\prod_{i=1}^N Y_i \cdot
<\L|(1+X_1F^{r_1}(\mu_1))\ldots (1+ X_N F^{r_N}(\mu_N))|\L>
\label{5.26b}
\bj
for any fundamental representation with highest weight vector $|\L>$
of the affine Lie algebra $A^{(1)}_{n}$. Therefore, the representations 
(\ref{1.1}) and (\ref{1.2}) coincide provided that 
\a
e^{\frac{\z-\z_0}{n+1}}&=&\prod_{i=1}^N Y_i
\label{5.27}
\b
We hope to go back to the proof of the above equation elsewhere.

\appendix
\section{Appendix: The Affine Lie Algebra $A^{(1)}_{n}$ in the Principal Gradation}

\renewcommand{\theequation}{\thesection.\arabic{equation}}
\setcounter{equation}{0}
\setcounter{subsection}{0}
\setcounter{footnote}{0}  

The purpose of this Appendix is to resume some basic facts from the theory
of the Lie algebras. We will focus our attention on the simple 
Lie algebras $A_n$ and on the untwisted affine Lie algebras $A^{(1)}_{n}$. More detailed exposition can be found in 
\cite{ Kac, OlS, Kne}. 

We recall that the Lie algebra $sl(n+1)$, \,\,
 $n \geqslant 1$ is the set of the traceless 
$(n+1) \times (n+1)$ matrices. Within the Cartan
classification they are known as $A_n$ Lie algebras. 
The Cartan subalgebra consists of all  traceless 
linear combinations of the diagonal elementary matrices
 $E^{ii}=|i><i|\,\,\,( i=1\ldots n+1)$. The root system 
can be embedded into the $(n+1)-$dimensional Euclidean 
space. Fixing an orthonormalized basis ${e_i}$, the 
roots are $\al_{ij}=e_i-e_j,~~ i\neq j\,\,\,( i,j=1,\, \ldots ,\,n+1)$. 
To each root one associates a step operator 
$E^{\al_{ij}}=E^{ij}=|i><j|\,\,\,( i \neq j)$ which is 
an eigenvector of the adjoint action of the Cartan 
subalgebra $\H$. As simple roots one chooses the 
vectors  $\al_{i}=e_i-e_{i+1},~~ i=1, ..., n$. The 
related step operators satisfy the commutation relations
\a
& &\left[H_\xi,E^{\pm\al_i}\right]=
\pm\al_i\cdot \xi \, E^{\pm\al_i}=\pm(\xi_i-
\xi_{i+1})E^{\pm\al_i}\0\\ 
& &\left[E^{\al_i},E^{-\al_j}\right]=
\delta_{ij}H_{\al_i}\0\\
& &H_\xi=\sum_i\xi_i|i><i|
\label{A.1}
\b

The generic step operators are obtained by successive
 commutators of $E^{\al_i}$ and their transposed $E^{-\al_i}$.
 
In the theory of the Lie algebras it is important to study their 
finite order inner automorphisms. The general
theory has been developed by Kac \cite{Kac} and reviewed in
\cite{gDS}. In this paper we shall only use a special inner
automorphism $\s$ of the simple Lie algebra $A_n$  of order
$n+1$ ($\s^{n+1}=1$). Before introducing it, we recall 
that the fundamental weights are
\a
& &\l_i=\sum_{k=1}^i e_k - \frac{i}{n+1}
\sum_{k=1}^{n+1}e_k\0\\
& &2\frac{\al_i \cdot \l_j}{\al_i\cdot \al_i}=
\delta_{ij},\,\,\,\,\,\ i,j=1,\,\ldots ,\, n
\label{A.2}
\b 
We also set $\rho=\sum_{i=1}^n \l_i$ and define \cite{Kac, DS} 
\a
\s(X)&=&S X S^{-1}\0\\
S&=&e^{ 2\pi i \frac {H_\rho}{n+1}}\label{A.3}
\b
for an arbitrary element $X$ of the Lie algebra $A_n$.
Note that in the defining representation, the element
$S$ which implements the automorphism  $\s$ takes 
the following form
\a
S&=& \o^{\frac{n}{2}} \sum_{k=1}^{n+1} \o^{1-k}
 E^{kk}\0\\
\o&=&e^{ \frac{ 2\pi i}{n+1}}~~~~\label{A.4}
\b 
From the commutation relations (\ref{A.1}) and
the above identity, one concludes that $\s$ acts 
trivially on the elements of $\H$ and as a multiplication 
by phase on the step operators:
\a
\sigma(H_\xi)&=&H_\xi\0\\
\sigma(E^{\al_{kl}})&=& \o^{\al_{kl}
\cdot \rho}E^{\al_{kl}}= \o^{l-k}E^{\al_{kl}}\0\\
\label{A.5}
\b
The Lie algebra $\G=A_n$ together with the automorphism $\s$
is a ${\ZZ}_{n+1}$ graded algebra
\a
\G&=&\oplus_{k\in{\ZZ}_{n+1}}\G_k\0 \\
\sigma(\G_k)&=&\o^k\G_k\0\\
\left[\G_k, \G_l \right]&\subseteq 
&\G_{k+l}\label{A.6}
\b
It is well known that for a given simple Lie
algebra $\G$, its Cartan subalgebra is fixed 
up to a conjugation by elements belonging to the 
corresponding Lie group. In particular, instead 
of $\H$, one can introduce an {\it alternative}
Cartan subalgebra $\H^{'}$, spanned on the 
mutually commuting generators:
\a
\E_i&=& \sum_{k=1}^{n+1-i} E^{kk+i}+
\sum_{k=1}^i E^{n+1+k-i k}=\0\\
&=&\sum_{k\in {\ZZ}_{n+1}}
 |k>< k+i|
\label{A.7}
\b
To show that the above elements actually generate
certain Cartan subalgebra, it suffices to note that
the matrix with entries
\a
& &\co_{ij}= \o^{(i-1)(j-1)}\0\\
& &\co^{-1}_{ij}=\frac{\o^{-(i-1)(j-1)}}{n+1}
\label{A.8}
\b
diagonalizes $\E_i$ (\ref{A.7}) 
\a
& &\co^{-1} \E_i \co= \sum_{k=1}^{n+1}
 \o^{i(k-1)} E^{kk}
\label{A.9}
\b
 It is worthwhile to note that the alternative 
Cartan subalgebra generators are eigenvectors 
of the inner automorphism $\s$
\a
& &\s (\E_i)=\o^i \E_i
\label{A.10}
\b
 For general simple Lie algebras, it is 
known  \cite{Kac, OlS} that the eigenvalues
of the corresponding automorphism $\s$, when 
restricted to the alternative Cartan algebra,
are in correspondence to the Betti numbers.
To complete the alternative basis of the Lie algebra $A_n$,
we introduce the generators
\a
& &F^{i}= \co E^{i+1i} \co^{-1}~~~~~, 
i=1, \ldots, n \0\\
& &\left[ \E_i , F^j\right]= ( \o^{ij}-1)F^j 
\label{A.11}
\b
Due to the grade decomposition (\ref{A.6}) and
the above commutation relations one gets
\a
& &F^i=\sum_{k\in {\ZZ}_{n+1}} F^i_k\0\\
& &\s (F^i_k) =\o^k F^i_k\0\\
& &\left[\E_i , F^j_k\right]= ( \o^{ij}-1)F^j_{k+i} 
\label{A.12}
\b
We thus end up with a  basis which is formed by
the alternative Cartan generators $\E_i$ (\ref{A.7})
and $F^i_k,,\,\,\, i=1,\ldots n,,\,\, k\in {\ZZ}_{n+1}$.
 Due to (\ref{A.8}) and (\ref{A.11}) the transformation 
which gives back the  Cartan--Weyl basis is
\a
& &E^{il}=\frac{\E_{l-i}}{n+1} + \sum_{r=1}^{n}
 \o^{r(1-i)} F_{l-i}^r,\,\,\,i<l \0\\
& &E^{il}=\frac{\E_{n+1+l-i}}{n+1} + \sum_{r=1}^{n}
 \o^{r(1-i)} F_{n+1+l-i}^r,\,\,\,i>l\0\\
& &E^{ii}-E^{n+1~n+1}=\sum_{r=1}^{n} \o^r(\o^{-ri}-1)F_{0}^r 
\label{A.13}
\b
The commutation relations in the
alternative basis are completed by  
 \a
& &\left[F^r_i , F^s_l\right]= \frac {\o^{sk}-\o^{rl}}{n+1}F^{r+s}_{k+l} 
\label{A.14}
\b
Introducing  notation
\a
&&\0\\
F^r_k&=&\left\{\begin{array}
{ccc}
F^r_k & 
r=1, \ldots, n;& k\in {\ZZ}_{n+1}\\

\\

 \frac{1}{n+1}\E_{k}
& r=0 & k=1, \ldots, n 
\end{array} \right. \label{A.15}
\b
we see that  the commutation relations in the alternative basis
assume the uniform expression (\ref{A.14}).

The $A_n$ Lie algebras are equipped with a 
nondegenerated invariant scalar product $(X,Y)=tr(X.Y)$.
The trace is taken in the defining representation.
In the alternative basis (\ref{A.15}) this scalar product
is given by
\a
(F^r_k ,F^s_l)=\frac{\o^{sk}}{n+1} \delta^{(n+1)}_{k+l,0}
\label{A.16}
\b
where  $\delta^{(n+1)}_{k,l}$ is the delta function in the cyclic
group ${\ZZ}_{n+1}$.
To treat integrable evolution equations, one has to extend the
classical Lie algebras by introducing the spectral (or loop)
parameter \cite{DS}. This wants to say that the Lax connection
belongs to the loop Lie algebra 
$\widetilde{\G}={\CC}[\l,\l^{-1}] \otimes \G$. In other words, 
the loop algebra is the set of the Laurent series with 
coefficients in the corresponding (classical) Lie algebra
 $\G$. Therefore $\widetilde{\G}$ is spanned on the elements 
$X_{k}=\l^k X,\,\, k\in {\ZZ},\,\, X\in \G$. The Lie
 bracket is
\a
\left[X_k , Y_l\right]&=& \left[X,Y\right]_{k+l}
\label{A.17} 
\b
Loop algebras $\widetilde{\G}$ possess central extension
\cite{Kac}, known as affine (or Kac--Moody) algebras 
$\hat{\G}=\widetilde{\G} \oplus {\CC}\hat{c} \oplus {\CC} \hat{d}$
\a
& &\left[X_k , Y_l\right]= \left[X,Y\right]_{k+l}
+\frac{k}{(n+1)}\hat{c}\, \delta_{k+l,0}\, (X,Y)\0\\
& &\left[\hat{d} , X_k\right]=k X_k\0\\
& &\left[\hat{c} , \hat{\G}\right]= 0
\label{A.18} 
\b
The normalization  factor which multiplies the central
element $\hat{c}$ is chosen for convenience. 
The derivation $\hat{d}$ can be used to 
define a ${\ZZ}$ gradation in $\hat{\G}$:
\a
\hat{\G}&=&\oplus_{k\in{\ZZ}}\hat{\G_k}\0\\
\left[ \hat{d} , \hat{\G_k}\right]&=&k \,\,\hat{\G_k}
\label{A.19}
\b
Since $\s$ is an automorphism of the underlying
classical Lie algebra $\G=A_n=sl(n+1)$, it is clear
that the commutators (\ref{A.18}) together with the restriction
\a
X(\o \l)=\s{X(\l)}\0\\
X(\l)=\sum_{l\in{\ZZ}} \l^l X_l\0\\
\s{X(\l)}= \sum_{l}\s (X_l)
\label{A.20}
\b
still define a Lie algebra. In the literature \cite{Kac, gDS, OlS} it 
is known as the affine  Lie algebra  $A^{(1)}_n$ in the principle 
gradation. 
Taking into account (\ref{A.14}),   (\ref{A.16}), (\ref{A.18}),
together with the above restrictions, we obtain the commutation
relations
\a
& &\left[F^r_k , F^s_l\right]=\frac{\o^{sk}-\o^{rl}}{n+1}F^{r+s}_{k+l}+ \frac{k\o^{rl}}{n+1}~\hat{c}~ \delta^{(n+1)}_{r+s,0}~\delta_{k+l,0}\0\\
& &~~~~~~~r,s\,\,\in {\ZZ}_{n+1};\,\,\,\,k,l \in {\ZZ}
\label{A.21} 
\b
Note that the generators $F^0_k$ with $k=0\, {\rm mod}(n+1)$ are not in 
the affine Lie algebra  $A_n^{(1)}$. We thus end up with the following alternative basis
\a
&&F^r_k,~~~~~~~r=1,\ldots ,\,n,\,\,\, k\in {\ZZ}\0\\
&&F^0_k,~~~~~~~ k\in {\ZZ},~~k\neq 0~{\rm mod}(n+1)\label{A.22}
\b
of the $A_n^{(1)}$ Lie algebra in the principal 
gradation. 
The generators $F^0_{k}, \,\,k \neq 0 {(\rm mod)}$ form a 
basis of Heisenberg subalgebra in the 
{\it principal} gradation \cite{Kac, DS,gDS,OlS}. 
In analogy with (\ref{A.15}),  we will also
use the following generators 
\a
& &\E_k=(n+1)F^0_k\0\\
& &\left[\E_k , \E_l\right]=(n+1)~k~ \delta_{k+l,0}\,\,\,\,
~k\neq 0\,(~{\rm mod}~n+1)
\label{A.23}
\b
From above commutation relations we conclude that 
$\E_{\pm k}$, $k\geq 1$ form an infinite set of noninteracting bosonic
oscillators.

\vskip 2cm

{\bf Acknowledgements} 
 
It is a pleasure to thank L. A. Ferreira and M. A. C. Kneipp 
for discussions on the subject. R. P. is also grateful to J. P. Zubelli for bringing to his attention the monograph \cite{Tomei}. Two of us, H. B. and
G. C. are supported  by CNPq--Brazil while R. P. is granted by 
FAPERJ--Rio de Janeiro. We deeply acknowledge both the foundations for 
the invaluable financial support.


\begin{thebibliography}{}
\bibitem{Raj} R. Rajaraman, {\it Solitons and
 Instantons}, North--Holland, Elsevier 1982.
\bibitem{FaKo} L. D. Faddeev and
 V. E. Korepin, Phys Rep. ${\bf 42}$, 1 (1978).
\bibitem{NS}  C. Nash and S. Sen, {\it Topology and Geometry
 for Physicits}, (Academic Press, Inc.)\\
M. Nakahara, {\it Geometry, Topology and Physics}, (Inst. Phys. Pub.)\\
M. G\"ockeler and T. Shucker, {\it Differential 
Geometry, Gauge Theories, and Gravity}, 
Cambridge Univ. Press.
\bibitem{Col} S. Coleman,  
Phys. Rev. ${\bf D11}$, 2088 (1975);\\
S. Mandelstam,  
Phys. Rev. ${\bf D11}$, 3026 (1975).
\bibitem{SW} N. Seiberg and E. Witten, 
Nucl. Phys. ${\bf B426}$, 19 (1994); 
Erratum  ${\bf B430}$, 485 (1994)\\
Nucl. Phys. ${\bf B431}$, 484 (1994).  
\bibitem{Ab} M. J. Ablowitz and H. Segur, 
{\it Solitons and the Inverse Scattering Method}, 
Siam, Philadelphia, 1981.\\
S. Novikov, S. Manakov, 
L. P. Pitaevsky and V. E. Zakharov, 
{\it  Theory of Solitons}, Consultants Bureau, 1984.\\
L. D. Faddeev and 
L. A. Takhtajan, {\it Hamiltonian Methods 
in the Theory of Solitons}, Springer, 1987.
\bibitem{Zam} A. B. Zamolodchikov and Al. B. Zamolodchikov,
 Ann. Phys.  ${\bf 120}$, 253 (1979).
\bibitem{T} M. Toda, Phys. Rep. ${\bf 18}$, 1 (1975).
\bibitem{OT} D. I. Olive and N. Turok,
 Nucl. Phys. ${\bf B215}$, 470 (1983); ${\bf 220}$, 491 (1983);
${\bf 257}$, 277 (1985), ${\bf 265}$, 469 (1986).
\bibitem{Kac}V. G. Kac, {\it Infinite 
dimensional Lie algebras}. Third edition.
Cambridge Univ. Press. 1990.  
\bibitem{DS} V. G. Drinfeld and V. Sokolov, 
Sov. Math. Dokl. ${\bf 23}$, 457 (1981).
\bibitem{gDS} N. Burroghs, M. de Groot, 
T. Hollowood and L. Miramontes,
Phys. Lett. ${\bf B277}$, 89 (1992);
Commun. Math. Phys. ${\bf 153}$, 187 (1993)\\
M. de Groot, T. Hollowood and L. Miramontes, 
Commun. Math. Phys. ${\bf 145}$, 57 (1992).
\bibitem{Fer} L. A. Ferreira, L. Miramontes 
and J. S. Guill\'en,  J. Math. Phys. ${\bf 38}$, 882 (1997).
\bibitem{Hol}T. Hollowood, Nucl. 
Phys. ${\bf B384}$, 523 (1992).
\bibitem{OlS} D. I. Olive, M. V. Saveliev and
 J. W. Underwood, Phys. Lett. 
${\bf B311}$, 117 (1993)\\
D. I. Olive, N. Turok and J. W. Underwood,
 Nucl. Phys. ${\bf B401}$, 663 (1993); ${\bf B409}$, 509 (1993)\\
M. A. C. Kneipp and D. I. Olive, Nucl. Phys. 
${\bf B408}$, 565 (1993).
\bibitem{Kne}M. A. C. Kneipp and
 D. I. Olive, Commun. Math. Phys.
 ${\bf 177}$, 561 (1996).
\bibitem{jap} E. Date, M. Jimbo, M. Kashiwara 
and T. Miwa, {\it Transformation Groups for
 Soliton Equations}, in: Non--linear integrable
 systems, eds M. Jimbo and T. Miwa 
( World Scientific, Singapore, 1983);
 Publ. RIMS, ${\bf 18}$, 1077 (1982)\\
\bibitem{STS} M. Semenov--Tian--Shansky,
 Publ. RIMS ${\bf 21}$, 1237 (1985);\\
{\it Poisson Lie Groups, Quantum Duality
Principle, and the Quantum Double},
 hep--th/9304042.
\bibitem{fr}O. Babelon and  D. Bernard, Phys.
 Lett. ${\bf B260}$, 81 (1991) ;\\
Commun. Math. Phys. ${\bf 149}$, 279 (1992).
\bibitem{BB} O. Babelon and D. Bernard,
 Int. J. Mod. Phys. ${\bf A8}$, 507 (1993).
\bibitem{Bel}  H. Belich and R. Paunov,
 J. Math. Phys. ${\bf 38}$, 4108 (1997).
\bibitem{Nid} M. Niedermaier,
 Commun. Math. Phys. ${\bf 160}$, 391 (1994).
\bibitem{Tomei}  R. Beals, P. Deift and C. Tomei,
 {\it Direst and Inverse Scattering on the line}, 
(Mathematical Surveys and Monographs, Nr 28).
\bibitem{FMc} H. Flaschka and D. W. McLaughlin, 
Prog. Theor. Phys, ${\bf 55}$, 438 (1976).
\bibitem{Bab} O. Babelon, D. Bernard and F. A. Smirnov, 
hep--th/9603010, and Commun. Math. Phys. ${\bf 182}$, 319 (1996);\\
Commun. Math. Phys. ${\bf 186}$, 601 (1997).
\bibitem{cat} O. Babelon and L. Bonora,
Phys. Lett. ${\bf B244}$, 220 (1990);\\
H. Aratyn, L. A. Ferreira, J. F. Gomes and
A. H. Zimerman, Phys. Lett. ${\bf B254}$, 372 (1991)372;\\
C. P. Constantinidis, L. A. Ferreira, J. F. Gomes and 
A. H. Zimerman, ibid. ${\bf B298}$, 88 (1993).
\bibitem{Date}E. Date, Osaka J. Math, 
${\bf 19}$, 125 (1982).
\bibitem{OlA} M. A. Olshanetsky and A. M. Perelomov, 
Phys Rep. ${\bf 71}$, 313 (1981).
\bibitem{body} O. Babelon and D. Bernard, 
Phys Lett. ${\bf B317}$, 363 (1993)\\
H. Braden and A. N. W. Hone, ibid. ${\bf B380}$, 296 (1996).
\bibitem{BJ} E. J. Beggs and P. R. Johnson,
Nucl. Phys. ${\bf B484}$, 653 (1997).
\bibitem{ZS} V. Zakharov and A. Shabat, 
Funct. Anal. Appl. ${\bf 13}$, 166 (1979).
\bibitem{GR} G. Cuba and R. Paunov, Phys.
 Lett. ${\bf B381}$, 255 (1996).
\bibitem{God} P. Goddard and D. Olive, Int J. Mod. Phys.
 ${\bf A1}$, 303 (1986)\\
P. Ginsparg, {\it Applied Conformal Field Theory}
Les Houches, Session XLIX, 1988, eds. E. Br\'ezin and J. Zinn--Justin.
\bibitem{Lep} J. Lepowsky and R. L. Wilson, 
 Commun. Math. Phys.  ${\bf 62}$, 43 (1978).
\bibitem{Kaz} V. G. Kac, D. A. Kazhdan, J. Lepowsky and R. L. 
Wilson, Adv. Math. ${\bf 42}$, 83 (1981).
\bibitem{Mac}N. J. Mackay and W. A. McGhee, Int. J. Mod. Phys.
${\bf A8}$, 2791 (1993).
\bibitem{Ver} E. Verlinde, Nucl. Phys. ${\bf B300}$[FS 22], 360 (1988).
\bibitem{Dub} B. Dubrovin,  Nucl. Phys. ${\bf B379}$, 627 (1992);
 Commun. Math. Phys ${\bf 152}$, 539 (1993).
\bibitem{SP}H. Aratyn, C. P. Constantinidis, L. A. Ferreira, J. F. Gomes
and A. H. Zimerman, Nucl. Phys. ${\bf B406}$, 727 (1993). 
\end{thebibliography}
\end{document}